\documentclass[twocolumn]{aastex62}

\usepackage{natbib}
\usepackage{amsmath}
\usepackage{textcomp}
\usepackage{graphicx}
\usepackage{ifpdf}
\usepackage{atbegshi}
\usepackage{amssymb}

\newcommand{\msun}{$M_{\odot}$}

\newcommand{\midline}{\, | \,}
\newcommand{\hi}{H$\,\textsc{i}$}
\def\tol[#1][#2][#3]{${#1}^{+{#2}}_{-{#3}}$}

\begin{document}

\title{SMASHing the LMC: A Tidally-induced Warp in the Outer LMC and a Large Scale Reddening Map}

\correspondingauthor{Yumi Choi}
\email{ymchoi@email.arizona.edu}

\author{Yumi Choi}
\affiliation{Steward Observatory, University of Arizona, 933 North Cherry Avenue, Tucson, AZ 85721, USA}
\affiliation{Department of Physics, Montana State University, P.O. Box 173840, Bozeman, MT 59717-3840, USA}

\author{David L. Nidever}
\affiliation{Department of Physics, Montana State University, P.O. Box 173840, Bozeman, MT 59717-3840, USA}
\affiliation{National Optical Astronomy Observatory, 950 North Cherry Ave, Tucson, AZ 85719, USA}

\author{Knut Olsen}
\affiliation{National Optical Astronomy Observatory, 950 North Cherry Ave, Tucson, AZ 85719, USA}

\author{Robert D. Blum}
\affiliation{National Optical Astronomy Observatory, 950 North Cherry Ave, Tucson, AZ 85719, USA}

\author{Gurtina Besla}
\affiliation{Steward Observatory, University of Arizona, 933 North Cherry Avenue, Tucson, AZ 85721, USA}

\author{Dennis Zaritsky}
\affiliation{Steward Observatory, University of Arizona, 933 North Cherry Avenue, Tucson, AZ 85721, USA}

\author{Roeland P. van der Marel}
\affiliation{Space Telescope Science Institute, 3700 San Martin Drive, Baltimore, MD 21218, USA}
\affiliation{Center for Astrophysical Sciences, Department of Physics \& Astronomy, Johns Hopkins University, Baltimore, MD 21218, USA}

\author{Eric F. Bell}
\affiliation{Department of Astronomy, University of Michigan, 1085 S. University Avenue, Ann Arbor, MI 48109-1107, USA}

\author{Carme Gallart}
\affiliation{Instituto de Astrof\'isica de Canarias, La Laguna, Tenerife, Spain}

\author{Maria-Rosa L. Cioni}
\affiliation{Leibniz-Institut f\"{u}r Astrophysics Potsdam (AIP), An der Sternwarte 16, D-14482 Potsdam, Germany}

\author{L. Clifton Johnson}
\affiliation{Department of Physics and Astronomy, Northwestern University, 2145 Sheridan Road, Evanston, IL 60208, USA}

\author{A. Katherina Vivas}
\affiliation{Cerro Tololo Inter-American Observatory, National Optical Astronomy Observatory, Casilla 603, La Serena, Chile}

\author{Abhijit Saha}
\affiliation{National Optical Astronomy Observatory, 950 North Cherry Ave, Tucson, AZ 85719, USA}

\author{Thomas J. L. de Boer}
\affiliation{Department of Physics, University of Surrey, Guildford, GU2 7XH, UK}

\author{Noelia E. D. No\"el}
\affiliation{Department of Physics, University of Surrey, Guildford, GU2 7XH, UK}

\author{Antonela Monachesi}
\affiliation{Instituto de Investigaci\'on Multidisciplinario en Ciencias y Tecnolog\'ia, Universidad de La Serena, Ra\'ul Bitr\'an 1305, La Serena, Chile}
\affiliation{Departamento de F\'isica y Astronom\'ia, Universidad de La Serena, Av. Juan Cisternas 1200 Norte, La Serena, Chile}

\author{Pol Massana}
\affiliation{Department of Physics, University of Surrey, Guildford, GU2 7XH, UK}

\author{Blair C. Conn}
\affiliation{Research School of Astronomy and Astrophysics, Australian National University, Canberra, ACT 2611, Australia}
\affiliation{Gemini Observatory, Recinto AURA, Colina El Pino s/n, La Serena, Chile}

\author{David Martinez-Delgado}
\affiliation{Astronomisches Rechen-Institut, Zentrum f\"ur Astronomie der Universit\"at Heidelberg,  M{\"o}nchhofstr. 12-14, D-69120 Heidelberg, Germany}

\author{Ricardo R. Mu\~noz}
\affiliation{Departamento de Astronom\'ia, Universidad de Chile, Camino del Observatorio 1515, Las Condes, Santiago, Chile}

\author{Guy S. Stringfellow}
\affiliation{Center for Astrophysics and Space Astronomy, University of Colorado, 389 UCB, Boulder, CO 80309-0389, USA}

\shortauthors{Choi et al.}

\begin{abstract}
We present a study of the three-dimensional (3D) structure of the Large Magellanic Cloud (LMC) using $\sim$2.2 million red clump (RC) stars selected from the Survey of the MAgellanic Stellar History. To correct for line-of-sight dust extinction, the intrinsic RC color and magnitude and their radial dependence are carefully measured by using internal nearly dust-free regions. These are then used to construct an accurate 2D reddening map (165~deg$^{2}$ area with $\sim$10\arcmin~resolution) of the LMC disk and the 3D spatial distribution of RC stars. An inclined disk model is fit to the 2D distance map, yielding a best-fit inclination angle $i$ = \tol[25.86][0.73][1.39] degrees with random errors of $\pm$0.19\degr\,and line-of-nodes position angle $\theta$ = \tol[149.23][6.43][8.35] degrees with random errors of $\pm$0.49\degr. These angles vary with galactic radius, indicating that the LMC disk is warped and twisted likely due to the repeated tidal interactions with the Small Magellanic Cloud (SMC). For the first time, our data reveal a significant warp in the southwestern part of the outer disk starting at $\rho\sim$7\degr\,that departs from the defined LMC plane up to $\sim$4~kpc toward the SMC, suggesting that it originated from a strong interaction with the SMC. In addition, the inner disk encompassing the off-centered bar appears to be tilted up to 5--15\degr\,relative to the rest of the LMC disk. These findings on the outer warp and the tilted bar are consistent with the predictions from the Besla et al. simulation of a recent direct collision with the SMC.  
\end{abstract}


\section{Introduction} \label{sec:intro}
The Large and Small Magellanic Clouds (LMC and SMC) are the largest satellites of the Milky Way (MW) and the closest interacting pair of dwarf galaxies.  For decades it was thought that the Magellanic Clouds (MCs) had completed many orbits around the MW and that the tidal forces of the MW had been the primary mechanism that created the Magellanic Stream \citep[e.g.,][]{gardiner96, yoshizawa03, connors04, connors06}. However, that picture changed dramatically in the last decade as $HST$ proper motions \citep{kallivayalil06b, kallivayalil06a, piatek08,kallivayali13} suggested that the MCs only recently fell into the MW potential \citep{besla07}, but that the LMC--SMC pair likely had been gravitationally bound to each other for more than several gigayears and possibly for a Hubble time \citep{besla07}. The pair has also likely experienced a recent direct collision \citep[e.g.,][]{olsen11,besla12,noel13,carrera17,zivick18}. It thus appears that the complex morphology of the LMC disk \citep[e.g.,][]{vandermarel01b,olsen02,besla16,mackey16} has more to do with the history of interactions with the SMC than with the MW. Indeed, the structure of the LMC seems to hold many keys to decoding the process of interaction between the MCs, and to their eventual fate as accreted satellites of the MW. 

Many more dwarf galaxies with smaller companions will be discovered with upcoming large telescopes in the next decade especially with LSST. They will shed light on galaxy merging mechanisms. The LMC--SMC pair, however, is a unique opportunity to closely witness an ongoing hierarchical merging event (i.e., the SMC interacting with the LMC and the MCs merging on to the MW) and to perform detailed studies of the building process of a large galaxy like the MW.  

The structure of the stellar populations and interstellar medium (ISM) across the LMC has been intensively studied, as the LMC is the closest \citep[49.9~kpc;][]{degrijs14} ``laboratory'' for the study of many astrophysical phenomena. The LMC has an inclined, slightly elongated, rotating, and star-forming disk \citep{kim98,vandermarel01b,vandermarel01a,vandermarel02,subramanian10} with a shallow stellar metallicity gradient \citep[e.g.,][]{carrera08,feast10,carrera11,piatti13,choudhury16,pieres16}, and one prominent spiral arm in the central region with an off-centered bar \citep{devaucouleurs72,zhao00,zaritsky04b,subramanian09}. Its depth along the line of sight is $\sim$5~kpc \citep{subramanian09,yanchulova17}, and the northern outer disk shows signs of disturbance in its stellar structure \citep{mackey16}. 

However, our understanding of the evolution of the LMC is still far from complete. This is mainly because most previous studies have heavily focused on the inner disk ($\lesssim$4\textdegree{} or $\lesssim$3.5~kpc at the LMC distance) containing the majority of the ongoing star formation and ISM. Although 2MASS \citep{skrutskie06} and DENIS \citep{epchtein97} observed the main body of the LMC, they still go out only to $\sim$7\degr\,\citep[e.g.,][]{vandermarel01b,vandermarel01a}. To understand the history of tidal interactions, it is necessary to explore the outer part of the LMC disk as well where (1) the potential is shallower and so more easily disturbed, and (2) older stellar populations are dominant and thus better trace the underlying disk structure. A good portion of the northern periphery of the LMC has been studied \citep[e.g.,][]{saha10,balbinot15,mackey16}, but the southern outskirt of the LMC has not been well explored.  

The Survey of the MAgellanic Stellar History (SMASH) is an NOAO community Dark Energy Camera \citep[DECam;][]{Flaugher15} survey of the Clouds mapping 480~deg$^2$ (distributed over $\sim$2400~deg$^2$ at $\sim$20\% filling factor) to $\sim$24th AB mag in $ugriz$ with the goal of identifying broadly distributed, low surface brightness stellar populations associated with the stellar halos and tidal debris of the Magellanic Clouds \citep{nidever17}. About one-third of the SMASH fields probe the main body of the LMC covering roughly 5\degr\,north and 10.5\degr\,south of the LMC center. We note that the southern periphery of the LMC is a novel region that has been barely explored in contrast to its northern periphery. 

In this study, we use red clump (RC) stars to map the dust reddening over 165~\MakeLowercase{deg}$^{2}$ and to explore the three-dimensional (3D) structure of the LMC disk. RC stars are abundant low-mass stars \citep[$\lesssim$2~\msun;][]{castellani00} in the core He-burning stage with intermediate age and moderately high metallicity, and constitute a prominent feature in the color-magnitude diagram (CMD). The RC stars occupy a well-defined and narrow region in the CMD since the stellar core mass at He ignition is nearly independent of their initial mass. This fundamental property results in a very limited range of effective temperatures and luminosities --- making RC stars essentially ``standard candles" --- and allows us to accurately measure the extinctions and distances for RC stars across the LMC \citep[and references therein]{girardi16}. These distances are then used to create a 3D map of the LMC to study and define its structure. Many studies have used RC stars as extinction and distance probes in the LMC \citep[e.g.,][]{girardi01,olsen02,koerwer09,subramanian09,haschke11,subramanian13,tatton13}. 

This paper is organized as follows. In Section~\ref{sec:data}, we briefly describe the SMASH survey and photometry data. In Section~\ref{sec:rc}, we present the RC selection, maps of RC star counts, median colors and magnitudes. Section~\ref{sec:intRC} presents the measured intrinsic RC color and magnitude in the LMC and their radial profiles. Then we present and validate our reddening map in Section~\ref{sec:ext}. We discuss the 3D structure of the LMC in Section~\ref{sec:3Dmap} as well as the physical implication of our findings and explore the dependence of the resulting 3D structure on the stellar population effect in Section~\ref{sec:discussion}. The main conclusions are summarized in Section~\ref{sec:summary}. 

\begin{figure*}
\centering
\includegraphics[width=16cm]{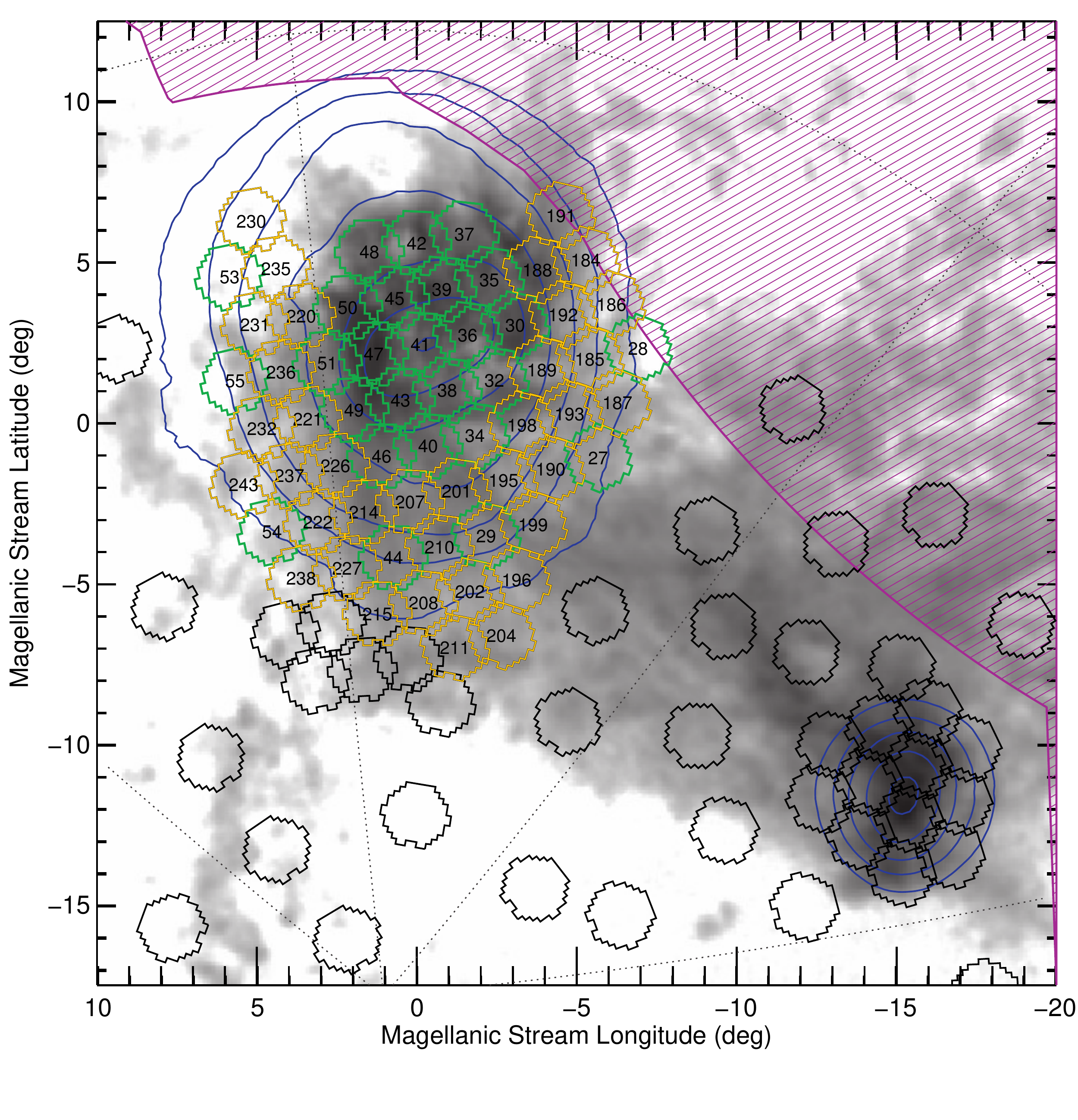}
\vspace{-1cm}
\caption{A portion of the entire SMASH survey covering the LMC and SMC. Grayscale shows the observed \hi\,column density of the area from the the Parkes Galactic All Sky Survey \citep[GASS;][]{mcclure-griffiths09}. Hexagons are the SMASH fields. The 62 fields used in this study are represented as green (deep exposure) and yellow (shallow exposure) hexagons with their field numbers. The underlying blue contours represent the 2MASS \citep{skrutskie06} red giant branch star counts. The purple shaded region represents the Dark Energy Survey footprint.  
\label{smash_fp}}
\end{figure*}

\section{Data and Photometry} \label{sec:data}
The SMASH survey builds on the technique first adopted by the Outer Limits Survey \citep{saha10}, namely, to use old main sequence (MS) stars as tracers to reveal the relics of the formation and past interactions of the Clouds.  With the large field of view of DECam, SMASH is able to use individual stars to probe down to surface brightnesses equivalent to $\Sigma_{g}=$ 35~mag~arcsec$^{-2}$ over a vast area.

Figure~\ref{smash_fp} shows the footprints of the SMASH survey fields in the region of the MCs over the \hi~gas map of \citet{mcclure-griffiths09}. \citet{nidever17} describes in detail the SMASH survey strategy, data reduction, and calibration as well as the first public data release containing $\sim$700~million measurements of $\sim$100~million objects in 61 deep and fully calibrated fields via the NOAO Data Lab\footnote{http://datalab.noao.edu/smash/smash.php}  \citep{fitzpatrick16}. In brief, the SMASH images are first reduced with the NOAO Community Pipeline \citep[CP;][]{valdes14} and then PSF photometry catalogs are generated with the DAOPHOT-based \citep{stetson87} PHOTRED pipeline\footnote{https://github.com/dnidever/PHOTRED}. These catalogs are calibrated using photometric transformation equations derived from standard star fields. We refer the readers to \citet{nidever17} for the detailed description of the data reduction and photometry. 

The photometric precision of the final SMASH catalogs is roughly 1.0\% ($u$), 0.7\% ($g$), 0.5\% ($r$), 0.8\% ($i$), and 0.5\% ($z$). The obtained calibration accuracies are 1.3\% ($u$), 1.3\% ($g$), 1.0\% ($r$), 1.2\% ($i$), and 1.3\% ($z$). The median 5$\sigma$ point source depths in the $ugriz$ bands are (23.9, 24.8, 24.5, 24.2, 23.5) mag, respectively, which is $\sim$2~mag deeper than SDSS \citep{york00} and $\sim$1.4~mag deeper than Pan-STARRS1 \citep{chambers16}. The astrometric precision is $\sim$15~mas, and accuracy is $\sim$2~mas with respect to the Gaia DR1 astrometric reference frame \citep{brown16}. 

\begin{figure*}
\centering
\includegraphics[width=18cm]{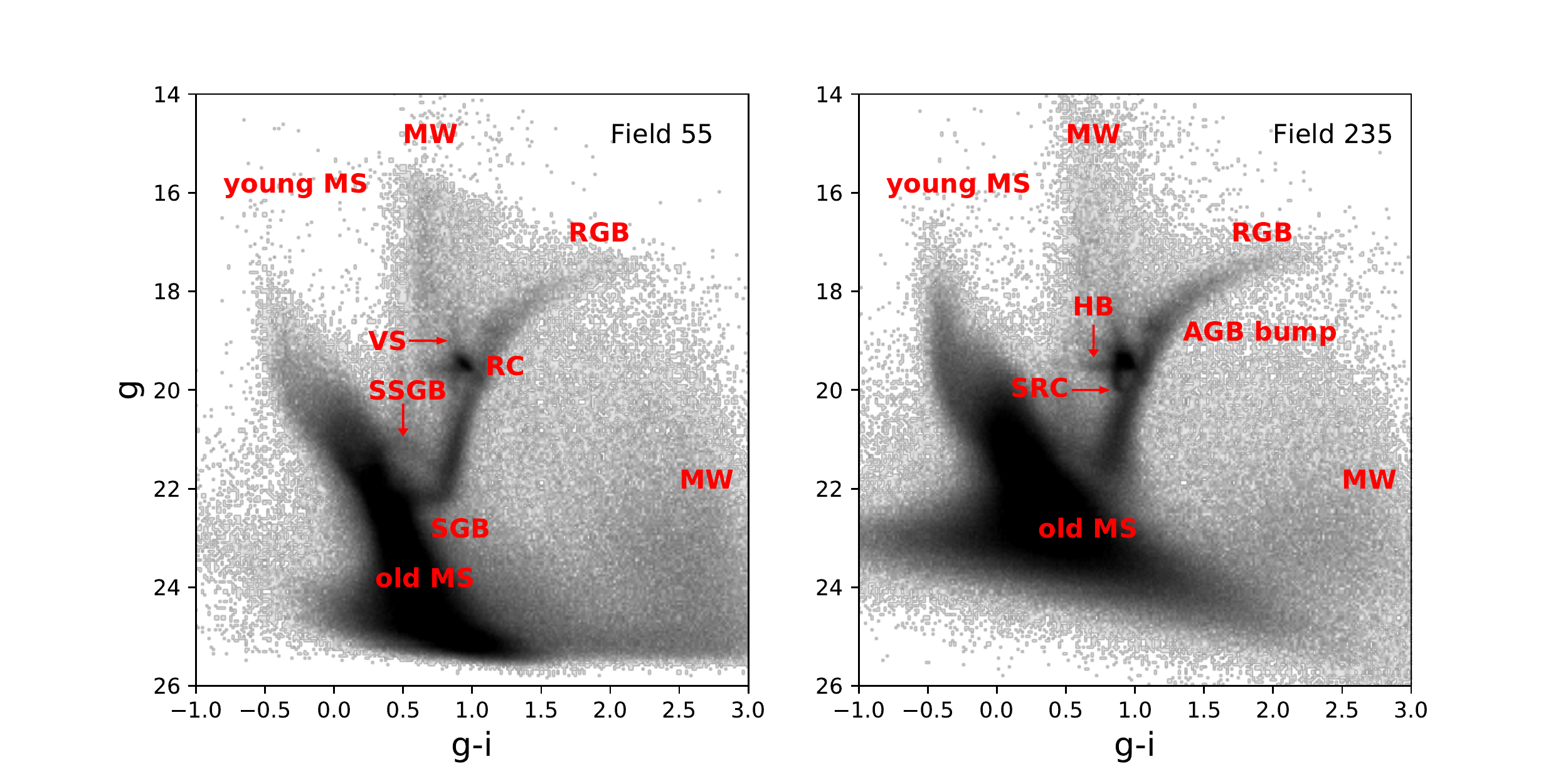}
\caption{Example CMDs from Fields 55 (left) and 235 (right). Each of them represents our deep exposure and shallow exposures, respectively. They show rich features containing information on the stellar populations in these fields. In addition to typical features such as a main sequence, a subgiant branch, a red giant branch, a red clump, and an asymptotic giant branch, there are unusual features as well. These include a secondary subgiant branch (SSGB), secondary red clump (SRC), a horizontal branch (HB), and vertical structure (VS; or blue loop). For clarity, we label the features in one of the CMDs with a clearer appearance of each feature.    
\label{exampleCMD}}
\end{figure*}

Out of the 197 SMASH survey fields, we use 62 observing fields\footnote{We note that only two (Field 44 and Field 55) of these 62 fields were included in the first data release.} that cover the LMC main body (up to $\sim$5\degr\, to the north and up to $\sim$10.5\degr\, to the south from the LMC center) and have noticeable LMC RC populations. Our data cover the outer portion of the southern LMC disk which is a region that has not yet been well studied. Forty of the outer LMC fields (all beyond a radius of 4.5\degr) do not have the deeper exposures and $u$-band data, and are therefore roughly $\sim$1.5 mag shallower than our standard SMASH fields. However, our artificial star tests (ASTs) show that the data are still sufficiently deep for studying the LMC RC stars.

To characterize the photometric completeness of our data, we perform ASTs. Initial ASTs are run on all the SMASH fields using a single CCD. Roughly 10,000 artificial stars are added to all of the exposures for a given field. The artificial stars uniformly sample the ($g-i$,$g$) CMD space ($-$1.0 $< g-i <$ 3.5 and 17.0 $< g <$ 27.0) and the spatial region on the sky covered by the images. For each image, the ($\alpha$,$\delta$) coordinates are converted to pixel positions using the WCS, and the calibrated magnitudes are converted into instrumental magnitudes using the photometric transformation equations for that field.  The artificial stars are then added to the image using DAOPHOT's ADDSTAR routine and the PSF of that image. Next, the images are processed with PHOTRED using the original settings and the new option for artificial stars, which skips the PSF generation and certain other steps. To determine the recovered artificial stars, the final catalog of objects are crossmatched with both the original science catalog and the input list of artificial stars. When a recovered object matched both an artificial star and an original object in position, the multiband photometry is used to find the best match. 

Completeness maps in ($g-i$,$g$) are generated for each field using the recovered artificial stars. The 50\% completeness at the RC color of $g-i$ = 0.8 for LMC fields beyond a radius of $\sim$5\degr\, is $g$$\sim$25~mag and decreases to $g$$\sim$22.5~mag in the innermost and most crowded LMC fields, but this is still three magnitudes deeper than the RC at $g$ = 19.3~mag. The completeness improves for redder colors; therefore, the photometry for reddened and fainter RC stars is also of good quality. The shallow LMC fields are $\sim$1.5 mag shallower than the normal SMASH fields, but the 50\% completeness for the RC color of $g$$\sim$23.5~mag is still over four magnitudes fainter than the RC. More complete ASTs using multiple mocks and covering all of the chips for a given field are currently in progress.

Figure~\ref{exampleCMD} presents example CMDs of deep and shallow exposure fields (Fields 55 and 235, respectively). Both CMDs show many detailed features associated with different stellar evolutionary stages. An MS covering many ages extends up to $i$$\sim$16~mag. Two subgiant branches are visible, one at an intermediate age ($\sim$3~Gyr) and a second at a much older age ($\sim$10~Gyr with $Z$ = 0.004). There is also a prominent red giant branch (RGB), asymptotic giant branch (AGB) bump, main RC, secondary RC (SRC; vertical feature fainter than the main RC), vertical structure or blue loop (VS; vertical feature brighter than the main RC), and a horizontal branch (HB). Both the SRC and VS are in the evolutionary stage of core He burning, but born with a slightly higher initial mass than the main RC \citep{girardi16}. The HB is the metal-poor equivalent of the RC, thus appearing on the bluer side of the main RC. The LMC RGB appears as a single broad band unlike in M31 and the SMC where the RGB is well split into visibly distinct sequences \citep{dalcanton15, yanchulova17}. A single RGB sequence suggests that stars and dust in the LMC are well mixed rather than having a thin dust layer within a stellar disk. Finally, there are foreground MW stars appearing as a fairly uniform ``sheet'' in the CMD with a pileup of stars between the LMC MS and RGB resulting from the blue edge of the MW MS turnoff stars.

\begin{figure}
\centering
\includegraphics[width=9cm]{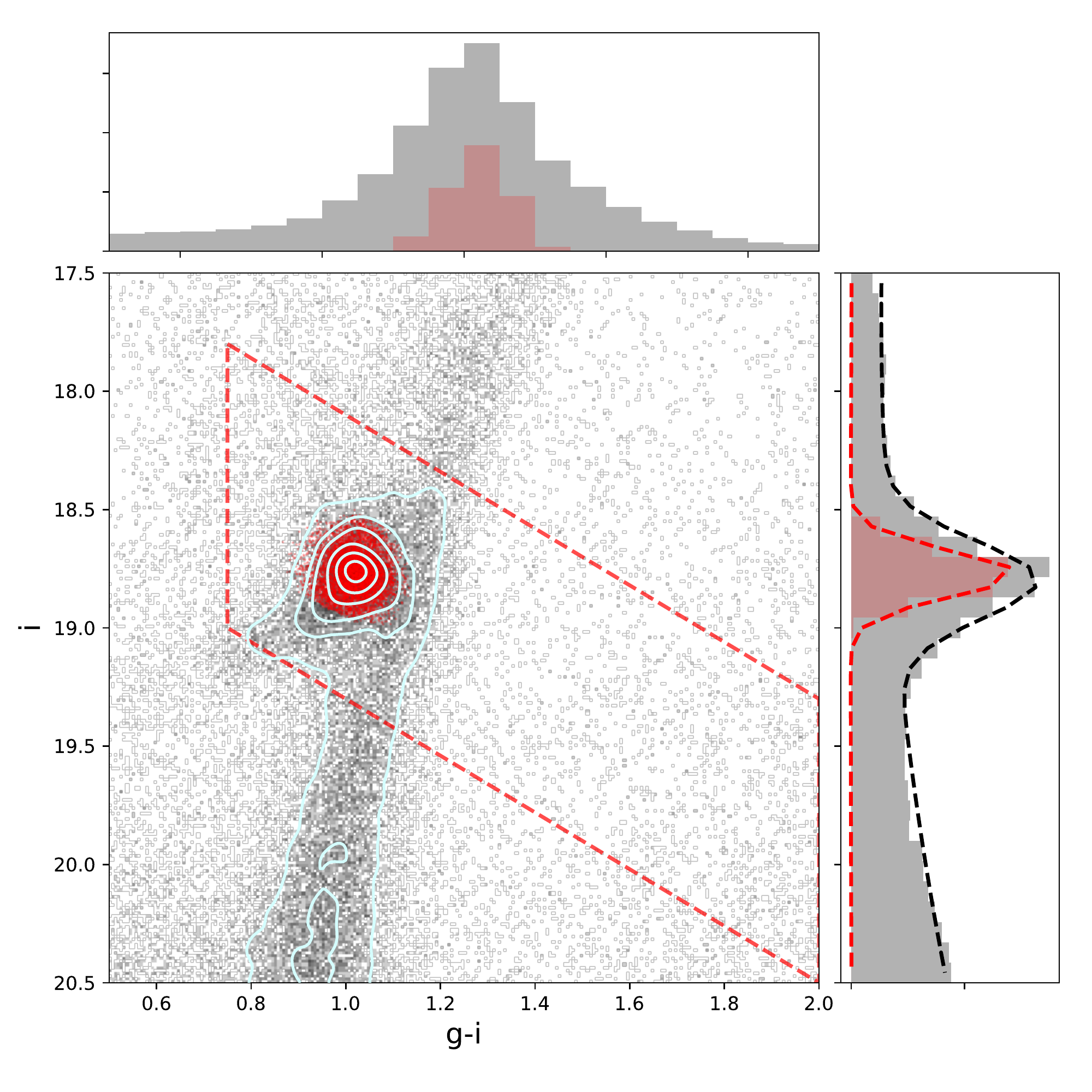}
\includegraphics[width=9cm]{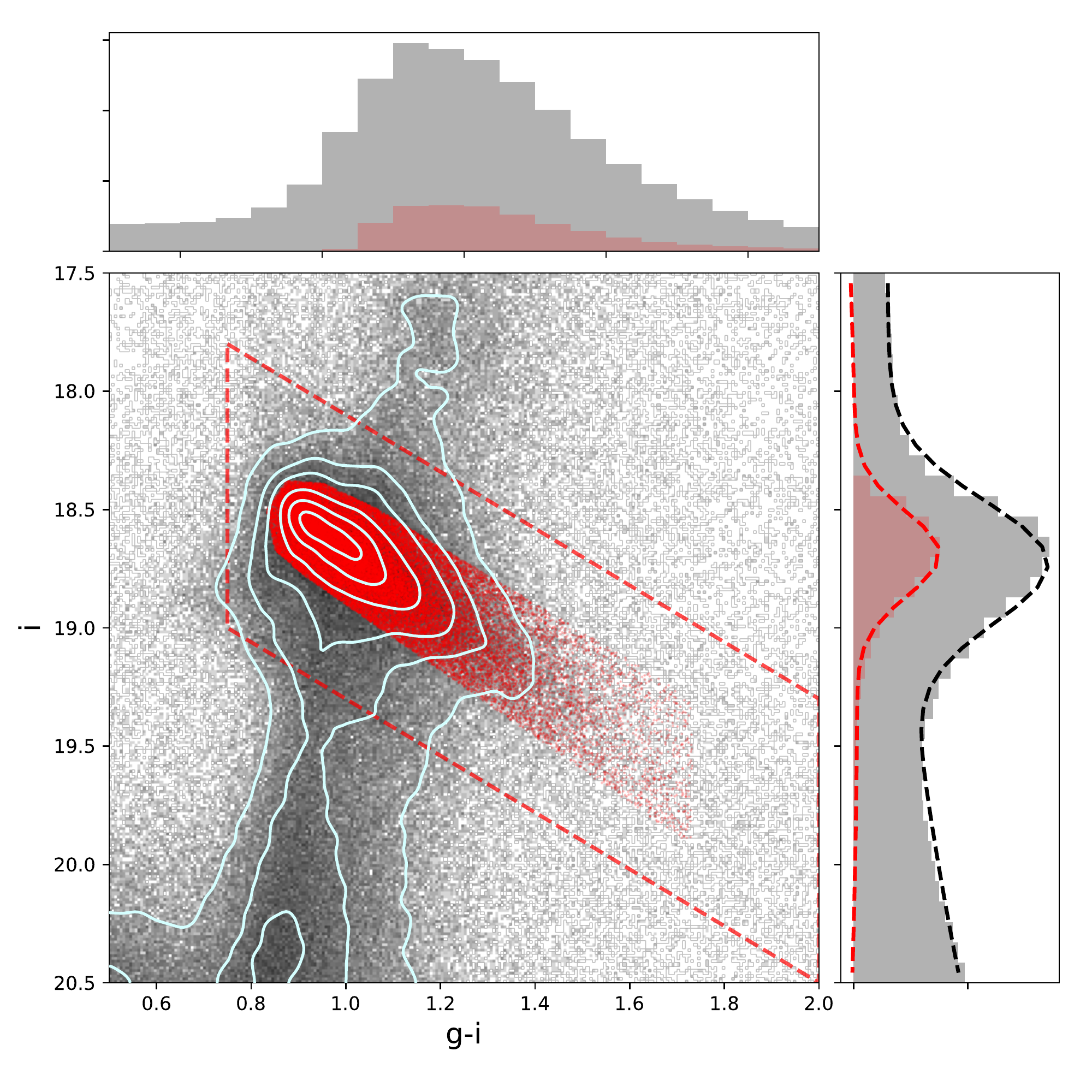}
\caption{Zoomed CMDs showing selected RC stars using our polygon (red dots) for Field 190 (top) as an example of a low-extinction field and 50 (bottom) as an example of a high-extinction field. The red dashed line shows an initial boundary to guide our selection. The slope roughly follows the reddening vector for $R_{V}$ = 3.4, which is an average $R_{V}$ for the LMC \citep{gordon03}. Our RC selection is conservative in that the selection minimizes contaminants of RGB, HB, SRC, and VS stars. For each CMD, we also present marginal histograms of the color and magnitude for both all stars (gray) and the selected RC stars (red) in the top and right-hand panels, respectively. A Gaussian plus second-order polynomial fit to each magnitude distribution is shown as a dashed line.
\label{selection}}
\end{figure}

\section{RC Stars in the LMC}\label{sec:rc}
The RC is one of the most prominent features in the CMD. In principle, the intrinsic color and brightness of the RC are confined to a narrow region in the CMD. Thus, the amount of dust extinction can be inferred from a shift in its observed color from the intrinsic color. Moreover, extinction-corrected magnitudes can then be used to determine the distances to individual RC stars, enabling the construction of a 3D map of a galaxy. In reality, however, the colors and magnitudes are affected by variations in the stellar populations (such as age and metallicity) due to spatial variation in star formation histories (SFHs) and metal enrichment history across the galaxy. In this study, we use the RC as extinction and distance probes of the LMC disk, with careful attention to effects other than extinction and distance that may influence our measurements. We will fully discuss the effect of variations in stellar populations in Section~\ref{sec:discussion}.  

\begin{figure}
\centering
\includegraphics[width=9cm]{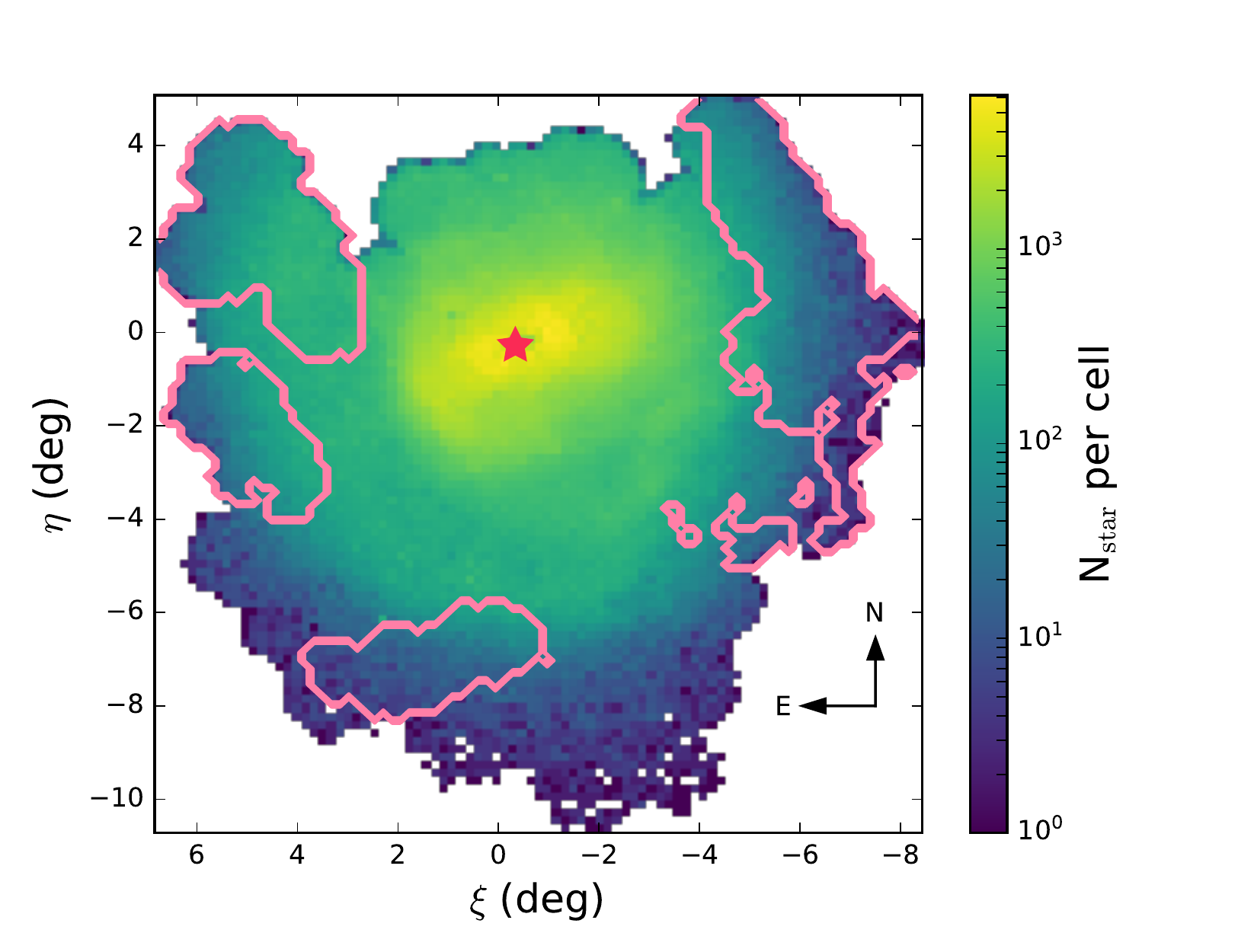}
\caption{RC star count map in stars per 10\arcmin$\times$10\arcmin~pixel. North is up and east is left. The red star indicates the center of the bar \citep{vandermarel01a}. The pink line outlines the region where the ``clean'' RC sample (i.e., RCs with zero or a negligible amount of internal dust) is distributed over (see Section~\ref{sec:int_col}). The electronic version of the RC count map is available as supplementary material to this paper. 
\label{starcount}}
\end{figure}

\begin{figure*}
\centering
\includegraphics[width=18cm]{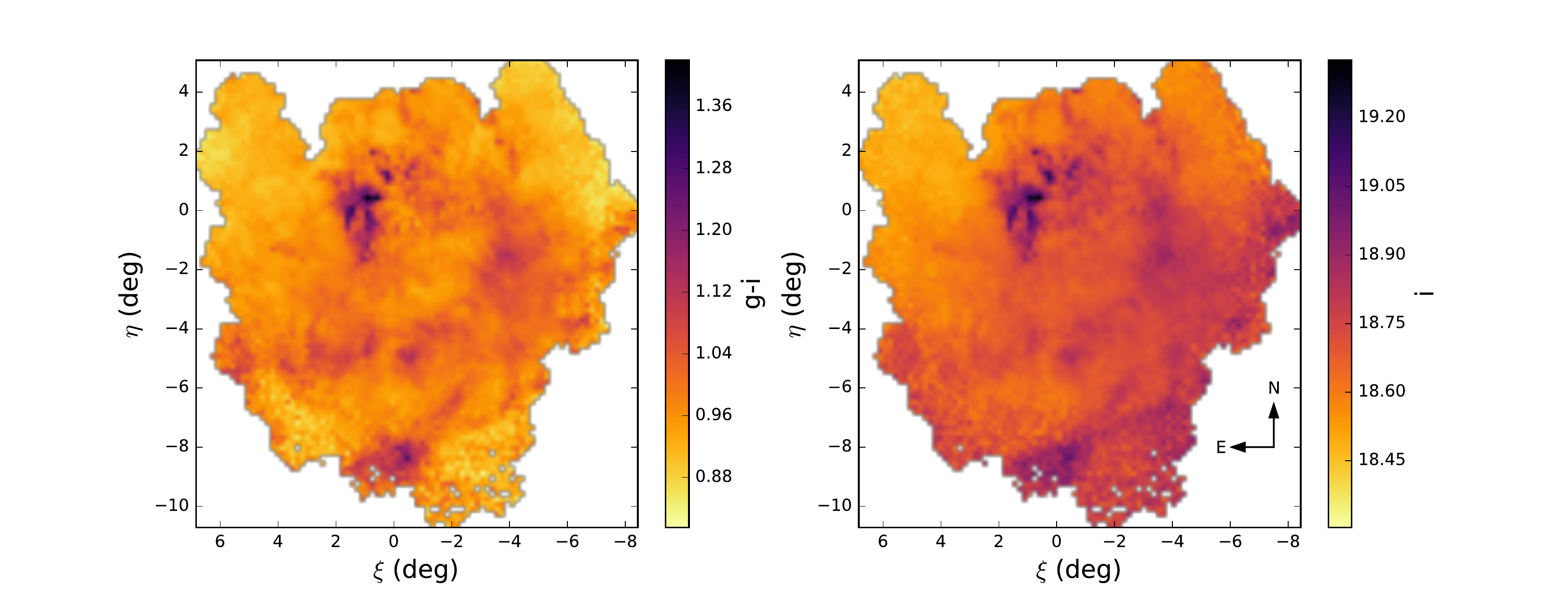}
\caption{Maps of the median $g-i$ color (left) and $i$-band magnitude (right) of the RC stars. The observed color map itself already traces well the dust emission in the inner disk as well as the MW galactic cirrus in the outer disk. The observed magnitude map shows an additional spatial gradient; the RC magnitude increases from northeast to southwest. These maps are available as supplementary material to this paper.
\label{obsmaps}}
\end{figure*}

\subsection{Selection of RC Stars}
In each SMASH field catalog, we first select point-like sources using the DAOPHOT ``chi'' and ``sharp''  parameters: chi$<$3 ($<$5 for Fields 36 and 41 in the center) and $\midline$sharp$\midline$$<$1 ($<$2 for Fields 36 and 41). Inside the LMC main body, contamination by background galaxies is negligible, especially within the magnitude range of interest in this study.

Figure~\ref{selection} describes our RC selection in each SMASH field. Selecting the RC stars based on their color and magnitude is straightforward. First, we define a large initial box (red dashed line) around the RC along the reddening vector and use it as a guiding boundary. The maximum color ($g-i=$ 2) is set in order to limit the contamination by MW foreground stars. To further minimize the contaminants in the RC selection, we carefully adjust the initial RC selection box by eye for each field, excluding stellar populations other than the main RC. The determined color range varies for each field mainly according to their relative dust extinction, while the slope is almost parallel to the reddening vector. Traditionally, other studies have used simple square selection boxes around the RC. For our CMDs, we found that such an approach led to an unacceptably high contamination of RGB stars and did not capture the entire RC in areas with high internal differential extinction. Once we select RC stars in each observing field, we remove duplicates in the overlapping regions among fields. The number of unique RC stars in our final catalog is $\sim$2.2~million.  

Figure~\ref{starcount} shows the number count map of the selected RC stars at 10\arcmin\,(150~pc) spatial resolution. We keep the same 10\arcmin$\times$10\arcmin\,spatial cell size for all subsequent analysis. We also conduct the same analysis at a much higher resolution of 40\arcsec\,(10~pc) and conclude that our main results are robust against the choice of spatial resolution. In general, the RC stars show an extended spatial distribution with smoothly decreasing number density with galactic radius, compared to the more concentrated and patchy distribution of young stars \citep[e.g.,][]{zaritsky04a} -- their star count map of young MS clearly shows a star-forming bar and one-armed spiral pattern. We note that high dust extinction makes our RC selection slightly incomplete only in the very central region ($\rho <$ 0.5\degr). In the rest of the disk with low crowding and low extinction, our RC selection is almost 100\% complete. The observed number count map itself already shows some interesting structures. We investigate the LMC disk structure for various stellar populations, such as RC and young MS, by modeling their observed star count map as a 2D projection of a tilted elliptical disk model in a separate paper \citep{choi18}.

\subsection{RGB Contamination}\label{sec:rgbcontamination}
Because the RGB usually overlaps with the RC in an observed optical CMD, it is impossible to completely remove RGB contaminants from the RC sample without extra information from spectroscopic data on surface gravity and effective temperature \citep[e.g.,][]{bovy14, nidever14}. Infrared (IR) color information could be helpful to distinguish the RC from RGB to some degree \cite[e.g.,][]{majewski11}. However, there are no publicly available IR survey data comparable with our optical data in terms of depth, photometric precision, and areal coverage even for the RC brightness level.

Fortunately, measuring extinction and distance does not depend significantly on the small fraction of RGB contaminants because they might induce only a small bias in the RC color and magnitude due to their similar color and magnitude compared to those of the RC, especially in each small 10\arcmin\,cell. Nevertheless, we estimate the fraction of possible RGB contaminants in our RC sample in each SMASH field. 
 
In Figure~\ref{selection}, we also show histograms of the color and magnitude for a given CMD. A Gaussian profile can describe the main RC distribution, and a second-order polynomial can describe the underlying distribution of contaminants including RGB stars \citep[e.g.,][]{stanek98, olsen02}. We fit the magnitude distribution of stars near the RC with a Gaussian profile combined with a second-order polynomial and find that the bright and faint end magnitudes of our selected RC stars are mostly within 2~$\sigma$ from the Gaussian mean, except for high-extinction fields. We calculate the RGB contamination fraction within our RC selection polygon based on the second-order polynomial terms. On average, the RGB contamination fraction in our sample is $\sim$10\%. If one uses a simple square box for the RC selection, the contamination fraction can be as high as $\sim$30--40\% by including additional populations on the bluer side (i.e., HB, SRC, and VS) as well as more RGB stars. This high contamination fraction would result in significant systematic errors in extinction and distance measurement by shifting the average color and magnitude from those of true RC stars.

\subsection{Observed Magnitude and Color Maps}
We first produce maps of the median observed color and magnitude at each 10\arcmin\,cell, from which we will derive the reddening and distance. While the mean RC color and magnitude have typically been calculated by fitting the distributions with a Gaussian profile combined with a second-order polynomial, we take the median magnitude and color for each cell. Due to our conservative RC selection, we found no significant contribution from second-order polynomial terms when describing the selected RC's magnitude distribution in each 10\arcmin\,cell. Thus, there is no significant difference between the geometric median and the Gaussian mean, except for cells at the edge of the maps that suffer from small number statistics where the geometric median provides more robust measurements. 

Figure~\ref{obsmaps} presents the maps of observed (i.e., undereddened) median $g-i$ color and $i$-band magnitude, which is less affected by dust than the bluer DECam filters and more complete than the $z$-band. The observed color map traces the dust emission in the inner disk \citep[see Fig.6 in][]{gordon14} as well as the MW galactic cirrus in the outer disk \citep{sfd98}. In addition, the observed magnitude map shows a spatial gradient with magnitude increasing from the northeast to the southwest. The LMC disk is known to be tilted in a way such that the northeast is closer (brighter) to us and the southwest is farther away (fainter) from us \citep[e.g.,][]{caldewll86,vandermarel01a,olsen02,mackey16}. We measure the inclination ($i$) and line-of-nodes position angle ($\theta$) of the LMC disk using the RC in Section~\ref{sec:3Dmap}.

\begin{figure*}
\centering
\includegraphics[width=18cm]{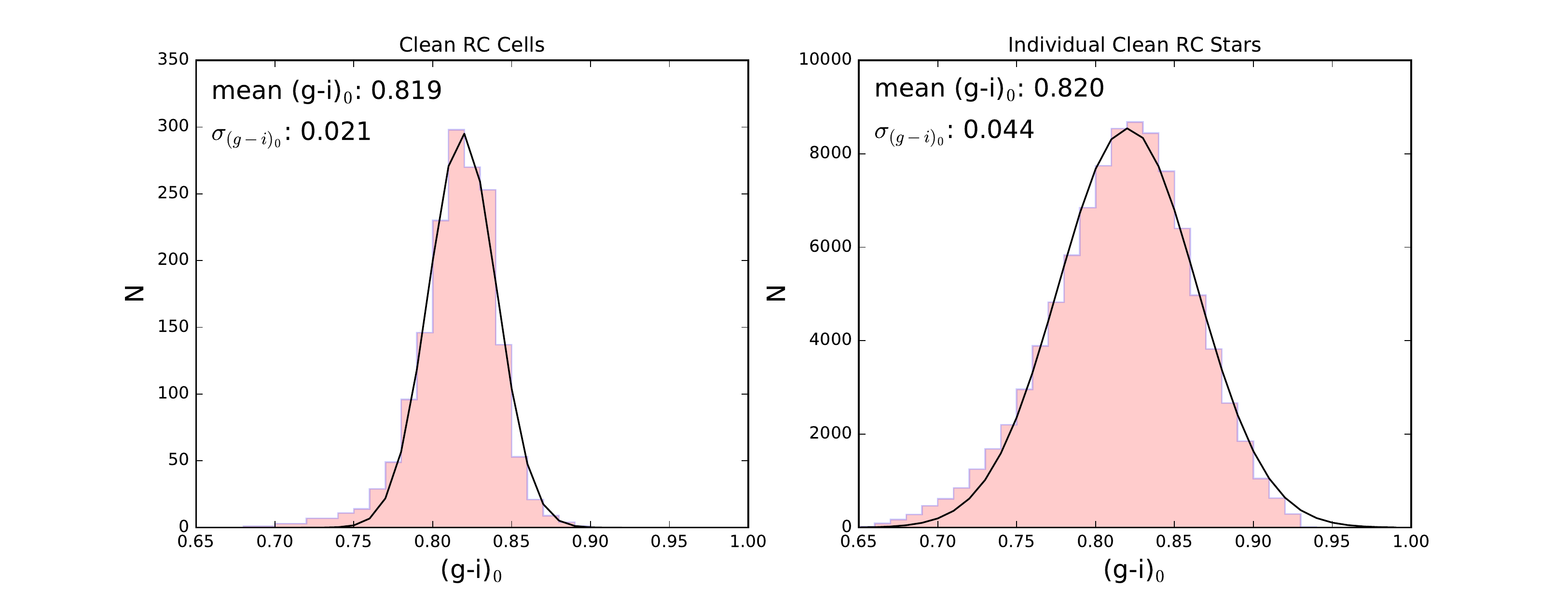}
\caption{The distribution of the intrinsic RC colors in the LMC. {\it Left:} Color distribution for all 10\arcmin\,cells in the clean RC region. It is well described as a Gaussian profile with a mean of 0.819 and $\sigma$ of 0.021. {\it Right:} Color distribution for individual clean RC stars (right). It is also well described as a Gaussian profile with the same mean of 0.820, but with the larger $\sigma$ of 0.044. This broader distribution for individual stars results from including both local and global stellar population effects, whereas the distribution for all cells reflects the global population effect only. 
\label{int_gi}}
\end{figure*}

\section{Intrinsic Properties of the RC}\label{sec:intRC}
For a given intrinsic color and magnitude of the RC, the observed color provides information about the line-of-sight dust (i.e., from reddening), while the observed magnitude is altered both by the dust and the RC distance. Although the intrinsic brightness and color of the RC are a function of stellar age and metallicity (i.e., stellar population), only a mild population gradient is expected in the LMC disk. This is because the LMC disk shares a relatively uniform SFHs for intermediate/old stellar populations \citep[e.g.,][]{harris09, weisz13, monteagudo18}. The metallicity gradient is also known to be mild across the LMC disk \citep[e.g.,][]{feast10}. Thus, one can expect no severe effects from stellar population differences across the LMC on the RC colors and magnitudes, except for the central region where the average age of the stellar population is younger. 

Although the LMC's population gradient is not expected to be severe, because we aim for a high-quality reddening map and an accurate distance measurement of the LMC disk, it is necessary to quantify the spatial variations in the intrinsic color and magnitude of the RC from differences in stellar populations in order to minimize and quantify systematic uncertainties in our 3D structure measurements of the LMC disk.

\subsection{Intrinsic Color of the Red Clump}\label{sec:int_col}
\subsubsection{Derived Intrinsic RC Color from a ``Clean'' RC Subsample}
To derive the intrinsic color distribution of the RC, we first construct a subsample of the RC in the regions with {\it no} or {\it negligible} internal dust. Because dust in front of these RC stars comes solely from the MW foreground, which acts as a dust screen, it is easy to break the degeneracy between the internal line-of-sight dust and the effect of stellar populations on their observed color. Using those RC stars enables us to derive their intrinsic colors after correcting for MW foreground dust. 

To construct the RC subsample, we start by selecting 18 SMASH fields that do not show noticeable reddened RC ``streaks" in their CMDs after correcting for MW foreground dust,\footnote{These fields are Field 27, 28, 44, 53, 55, 184, 185, 186, 187, 190, 191, 193, 220, 227, 230, 232, 235, and 238.} indicating zero or negligible amount of internal dust. It is highly unlikely that all RC stars in one SMASH field (a field of view of $\sim$3~deg$^{2}$) experience the same amount of internal dust, and thus do not develop a streak. For the MW dust correction in these 18 observing fields, we use the \citet[hereafter SFD98]{sfd98} reddening map, which was derived from infrared dust emission. To convert SFD98 $E(B - V)$ to $E(g - i)$, we use updated reddening coefficients\footnote{Reddening coefficients for other $R_{V}$ values are kindly provided by Keith Bechtol through private communication.} with $R_{V}$ = 3.1 for the DECam standard bandpasses that reflect Schlafly \& Finkbeiner's (2011) calibration adjustment \citep{abbott18}. Although the above 18 fields are located outside the regions where the emission from the LMC dominates FIR emission, a tiny fraction of 3 fields out of 18 overlaps with the inner region of the LMC where the internal dust becomes dominant. We first exclude RC stars residing in these overlapping regions. From the remaining RC stars in the 18 observing fields, we further exclude stars with SFD98 $E(B - V)$ values larger than 0.099 (outside 1$\sigma$ of the mean) to minimize potential contamination by RC stars that might be slightly reddened by a small amount of internal dust without developing a clear streak feature in a CMD. We designate this internal-dust-free RC sample as the ``clean'' RC sample, which accounts for $\sim$4\% of our RC star catalog. The clean RC sample is distributed over the regions that are outlined in pink in Figure~\ref{starcount}. The median SFD98 $E(B - V)$ value of the clean RC sample is 0.065~mag, which is consistent with a typical reddening value of 0.075~mag within uncertainty toward the LMC that was estimated based on the median dust emission arising from surrounding annuli \citep{sfd98}. 

Using only the clean RC sample, we compute the median intrinsic color for each 10\arcmin\,cell over the clean RC region outlined by the pink line in Figure~\ref{starcount}. In Figure~\ref{int_gi}, we present their median intrinsic color distribution (left) as well as the intrinsic color distribution of individual clean RC stars (right) for completeness. Both distributions are well described as a Gaussian profile with an identical mean $(g-i)_0$ of $\sim$0.82, but with different widths, with narrower width for the cells' median color distribution, as expected. The median RC color distribution of 10\arcmin\,cells reflects the cell-to-cell variation in SFH and chemical enrichment (i.e., global variations in stellar populations). The color distribution of individual clean RC stars, on the other hand, reflects both the full SFH and chemical enrichment history sampled in each cell (i.e., the local distribution of stellar populations) and their spatial variation. Almost all of the 10\arcmin\,cells exhibit standard deviations larger than $\sim$0.02 in their intrinsic color distribution of individual clean RC stars. The fact that the median color distribution of 10\arcmin\,cells in the clean RC region is narrower than the color distribution of individual clean RC stars per cell suggests that global variations in stellar populations across the LMC are moderate, at least for those populations that produce RC stars. 

\begin{figure}
\centering
\includegraphics[width=9cm]{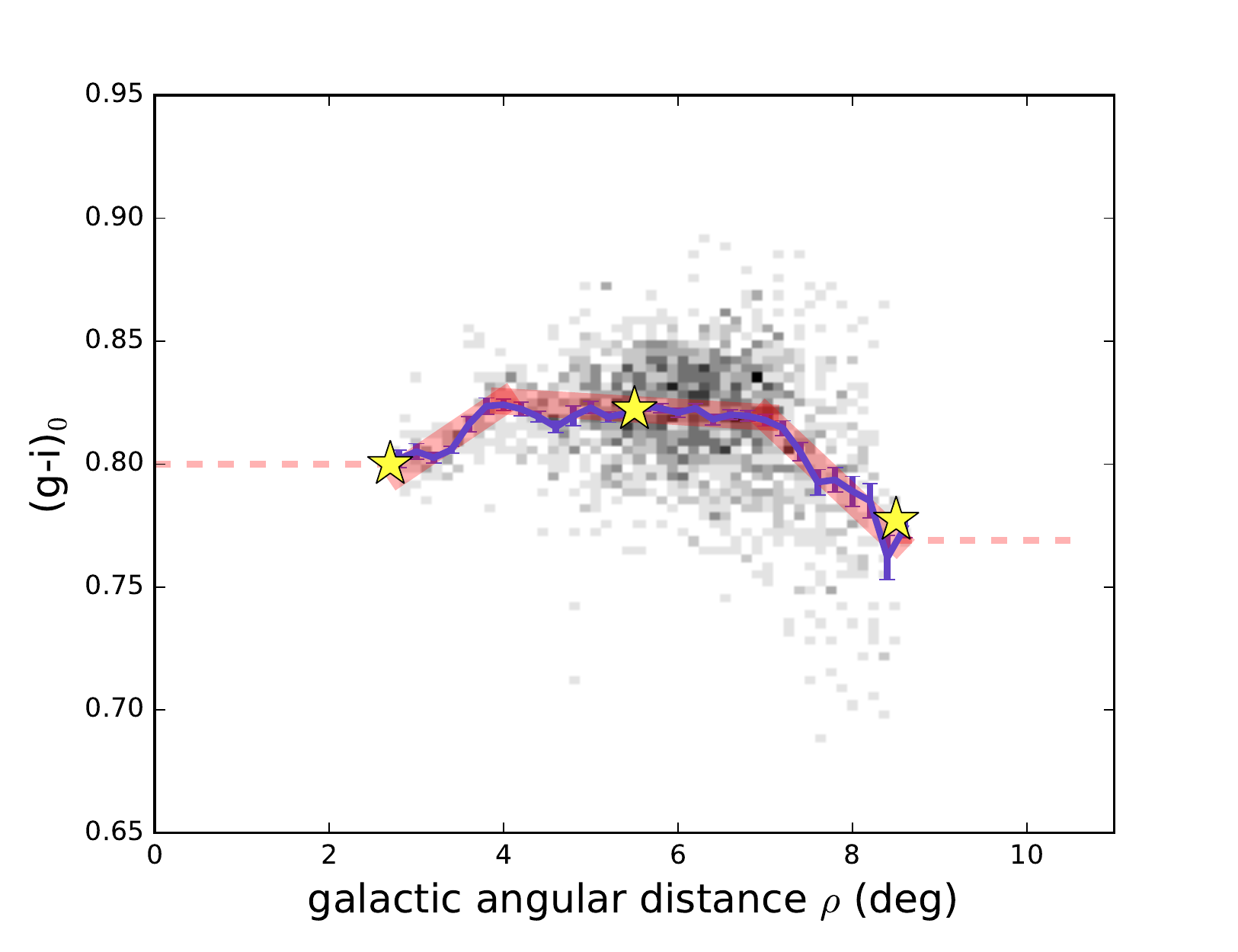}
\caption{Radial dependence of the intrinsic RC color constructed from the ``clean'' (i.e., zero or negligible internal dust) RC sample. The $\rho$ denotes galactic distance from the LMC center in degrees. The blue line shows the mean intrinsic color with an error bar in each 0.2\degr\,radial bin between 2.7\degr\,$< \rho <$ 8.5\degr. This limited radial range is due to the restricted radial coverage of the clean RC sample. The red shaded line is the best fit to the underlying grayscale data, and it is our default radial profile for the intrinsic RC color between 2.7\degr\,$< \rho <$ 8.5\degr. We measure an almost zero slope between 4\degr\, $< \rho <$ 7\degr\, with a constant color of 0.822, a slope of 0.024~dex~deg$^{-1}$ between 2.7\degr\,$< \rho <$ 4\degr, and a slope of --0.033~dex~deg$^{-1}$ between 7\degr\,$< \rho <$ 8.5\degr. The red dashed line shows our default radial profile for the intrinsic RC color at $\rho <$ 2.7\degr\,and $\rho >$ 8.5\degr. The yellow stars present the predicted $g-i$ color from the known metallicity gradient and the age-metallicity relation for the LMC disk at $\rho$ = 2.7\degr, 5.5\degr, and 8.5\degr.   
\label{rad_col}}
\end{figure}

\begin{figure*}
\centering
\includegraphics[width=18cm]{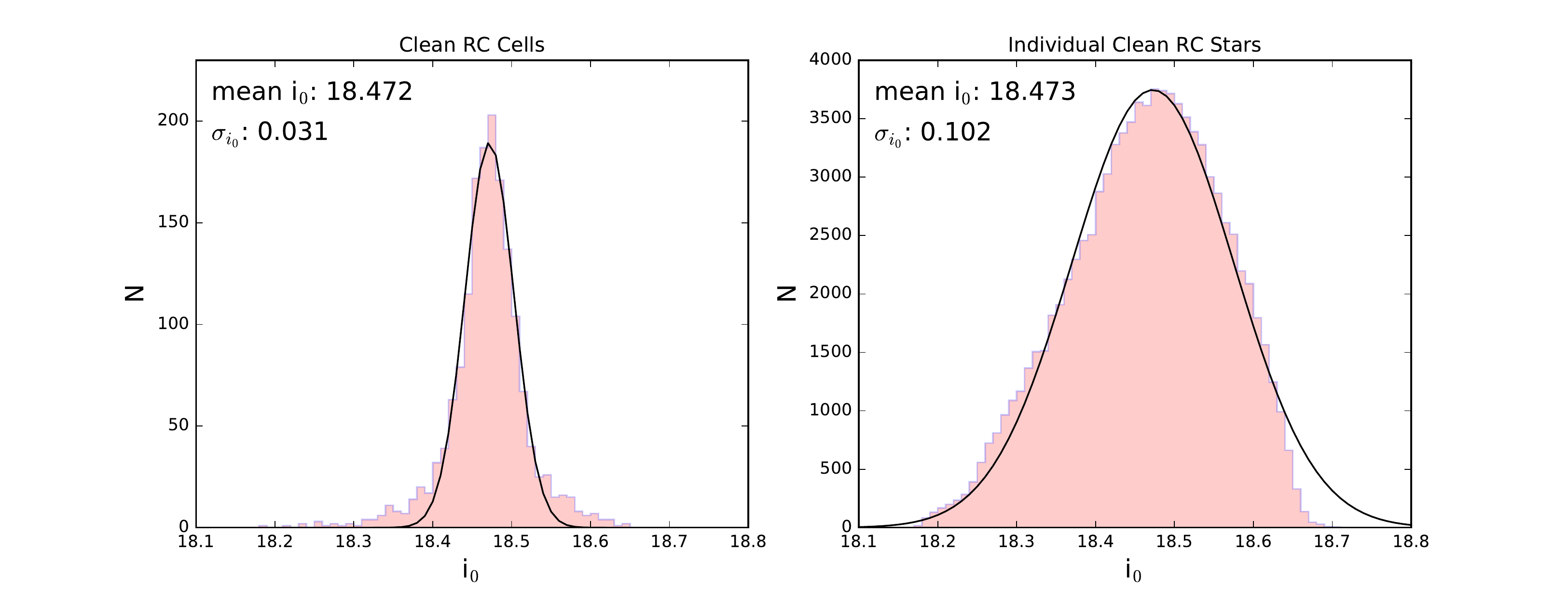}
\caption{Same as Figure~\ref{int_gi}, but for the intrinsic $i$-band magnitude that are derived from the clean RC sample after correcting for dust extinction and inclination effect. The intrinsic $i$-band magnitude distribution for all 10\arcmin\,cells is shown on the left, and that for individual clean RC stars is shown on the right. They have the same mean $i$-band magnitude of $\sim$18.47~mag, but the distribution for individual stars is more than three times broader than that of the 10\arcmin\,cells, again due to the contribution from both local and global population effects. The $\sigma$ of 0.031~mag for the cells' median $i$-band magnitude distribution, which represents the global population effect, is consistent with the amount of population effect in the $I$-band across the LMC fields \citep{girardi01}.
\label{int_i}}
\end{figure*}

\begin{figure}
\centering
\includegraphics[width=9cm]{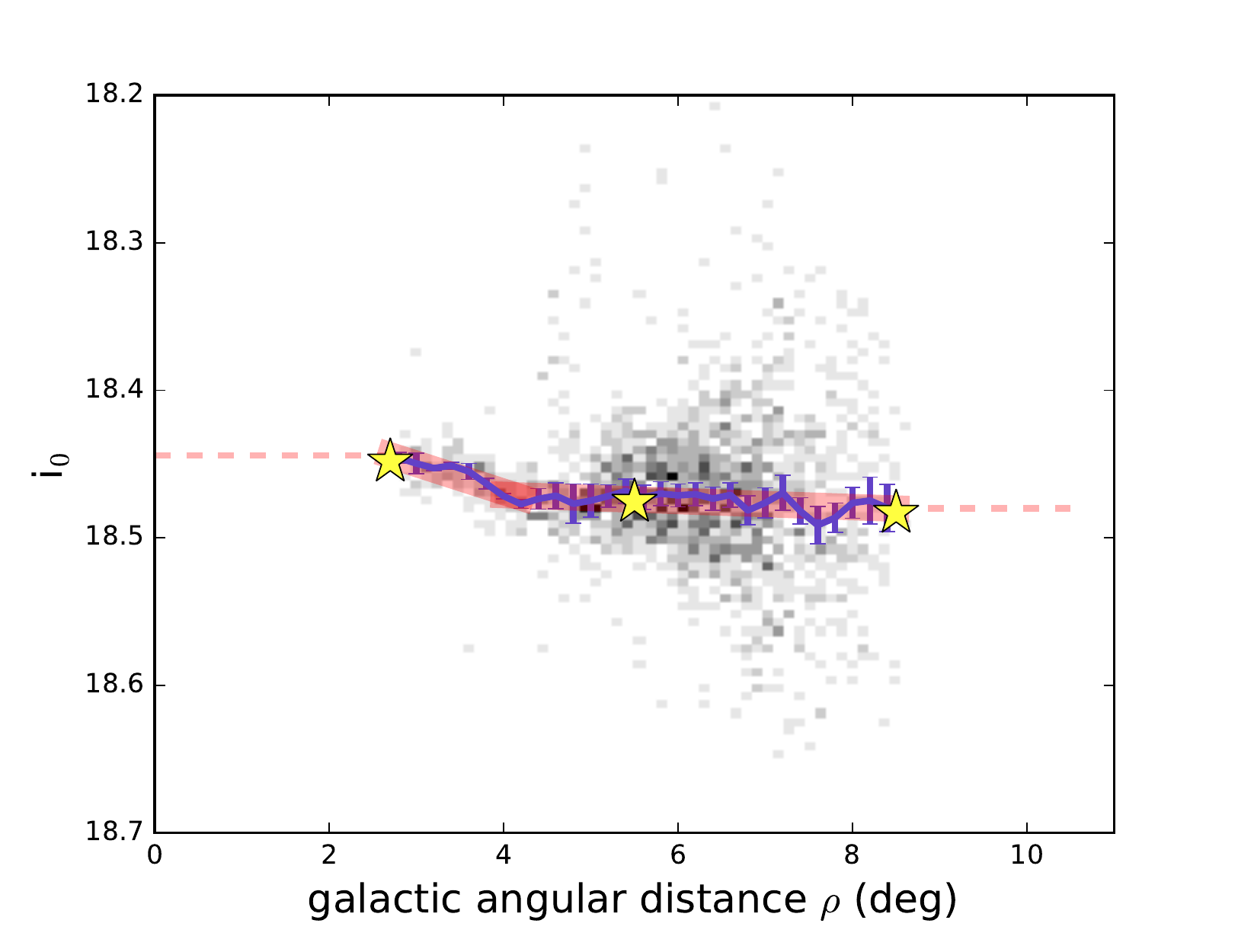}
\caption{Same as Figure~\ref{rad_col}, but for the intrinsic RC $i$-band magnitude constructed from the clean RC sample. The yellow stars show the predicted model RC magnitudes corresponding to the same metallicities and ages that are used to predict the model RC colors shown in Figure~\ref{rad_col}. The red shaded and dashed lines present our default radial profile of the intrinsic RC magnitude to derive the distance between 2.7\degr\,$< \rho <$ 8.5\degr and outside this radial range, respectively.  
\label{rad_mag}}
\end{figure}

\subsubsection{Intrinsic Color Radial Profile}\label{sec:col_rad}
In Figure~\ref{rad_col}, we investigate a radial profile of the intrinsic color using the clean RC sample. Because the clean RC sample consists of stars selected in the restricted area (see Figure~\ref{starcount}), it covers a limited range of galactic radius ($\rho$ in degrees) from $\sim$2.7\degr\,to $\sim$8.5\degr. The grayscale shows the distribution of the intrinsic colors of individual cells as a function of galactic radius from the LMC center, and the blue solid line presents a mean color with an associated error in each 0.2\degr\,width radial bin. The red shaded line shows the best fit to the gray data characterizing the overall radial profile of the observed intrinsic RC color. We find no statistically meaningful radial dependence of the intrinsic RC color between 4\degr\,and 7\degr. In this radial range, there is basically no change in the mean color with galactic radius, but there is a larger scatter around the median colors at larger radius. On the other hand, the intrinsic RC color tends to be slightly bluer both toward the center, with a slope of 0.024~dex~deg$^{-1}$, and the outer galaxy, with a slope of --0.033~dex~deg$^{-1}$.  

This RC color radial trend can be interpreted as a result of more metal-rich, but younger populations in the inner region and older, but more metal-poor populations in the outer region. An outside-in formation has been suggested for the LMC, supported by the observational evidence that old and metal-poor stars are preferentially found in the outer disk, whereas young and metal-rich stars are found in the inner disk \citep[e.g.,][]{gallart08,meschin14}. Although deriving full SFHs can provide a complete picture for the intrinsic RC properties across the LMC disk, this is beyond the scope of this paper. In a future paper, we will map the full SFHs using the SMASH data (C. Gallart et al., in preparation). 

To verify whether the detected radial color trend is physically reasonable, we conduct a simple comparison of the intrinsic RC colors with PARSEC isochrones \citep{bressan12,marigo17} by using previously reported metallicity gradients and age-metallicity relations (AMRs) in the LMC disk \citep{carrera08,carrera11,piatti13,choudhury16,pieres16} as a guide. These previous studies reported a shallow metallicity gradient with the mean \mbox{[Fe/H]} decreasing from --0.4 in the star-forming bar region to --0.8 to --1.0 in the outer region ($\geq$7\degr). There is no solid metallicity measurement for stars in the southern regions of the LMC outer disk yet. However, our CMDs can provide a loose constraint on it. For example, low-metallicity populations develop an extended horizontal branch feature in the CMD. We found no prominent extended HB features in any of our CMDs across the LMC disk, indicating that the minimum metallicity of the dominant stellar populations cannot be lower than [Fe/H]$\simeq$--2 even in the outer disk. Therefore, we set the minimum average \mbox{[Fe/H]} as --1.0, which is the minimum average \mbox{[Fe/H]} in the northern regions of the the LMC outer disk \citep{carrera08}.

We first compute the mean absolute $g$- and $i$-band magnitude as well as the mean $g-i$ color only for core He-burning stars from an isochrone at a given age and metallicity by following \citet[see their Equations 3 and 4]{girardi01}. The age is in steps of $\Delta\,log(age)$ = 0.5 and the metallicity ranges from \mbox{[Fe/H]} = --0.4 to --1.0 in steps of 0.1~dex. We use the Chabrier initial mass function to be consistent with the PARSEC isochrones used in this study. The calculated model RC colors and magnitudes show the same behavior presented in \citet{girardi01} and \citet{girardi16}. In short, the RC stars older than $\sim$2~Gyr behave in a simple way -- they become fainter and slightly bluer (redder) as the age (metallicity) increases. However, the colors and magnitudes for the RC stars younger than $\sim$1.5~Gyr vary rapidly with age and behave differently for different metallicities. We refer the readers to \citet{girardi01} and \citet{girardi16} for the detailed discussion about population effects on the RC photometric properties.  

We compare these calculated model mean colors for a given age and metallicity with the clean RC sample's intrinsic colors in each radial range: (i) 2.7\degr\,to 4\degr, (ii) 4\degr\,to 7\degr, and (iii) 7\degr\,to 8.5\degr. For these three radial bins, we adopt representative \mbox{[Fe/H]} values of --0.5, --0.7, and --0.8, respectively, which reflect the reported metallicity gradient in the LMC \citep{carrera08,carrera11,piatti13,pieres16}. Combining these three representative metallicities with the known AMR \citep{carrera08,piatti13}, we derive stellar ages of $\sim$1.6~Gyr, $\sim$5.6~Gyr, and $\sim$6.3~Gyr for these radial bins by finding models that best match our observed intrinsic colors. The best model colors in each radial bin are marked as yellow stars in Figure~\ref{rad_col}.

For the innermost radial bin (0\degr\,to 2.7\degr), we adopt the intrinsic RC color at $\rho =$ 2.7\degr\,as a default color, and for the outermost radial bin (8.5\degr\,to 10.5\degr), we adopt the intrinsic RC color at $\rho =$ 8.5\degr\,as a default color rather than introducing additional color gradients in these radial bins.

We, however, can attempt to infer physically reasonable intrinsic RC colors both inside $\rho =$ 2.7\degr\,and outside $\rho =$ 8.5\degr. For example, we can rule out the possibility of extrapolating the inner color gradient ($\sim$0.024~dex~deg$^{-1}$) to the innermost radial bin. This is because the metallicity gradient in the central bar region has been known to be shallow with a mean \mbox{[Fe/H]} of about --0.4~dex \citep[e.g.,][]{carrera08,carrera11,choudhury16}. If we extrapolate the inner color gradient of $\sim$0.024~dex~deg$^{-1}$ to the very center of the LMC, this predicts a $g-i$ color of $\sim$0.73. However, none of the metal-rich (\mbox{[Fe/H]} $>$ --0.5) isochrones have colors that blue and, in fact, all of the extrapolated colors for $\rho <$ 2.7\degr\,are too blue (which are bluer than $g-i \simeq$ 0.8) for the metal-rich isochrones, even with very young ages. Therefore, this color extrapolation to smaller radii does not appear physically reasonable. 

In Section~\ref{sec:popeffect}, we will investigate physically reasonable intrinsic RC colors in the innermost and outermost radial bins, and fully discuss the dependence of the 3D structure determination on the choice of the intrinsic RC colors.

\begin{figure*}
\centering
\includegraphics[width=18cm]{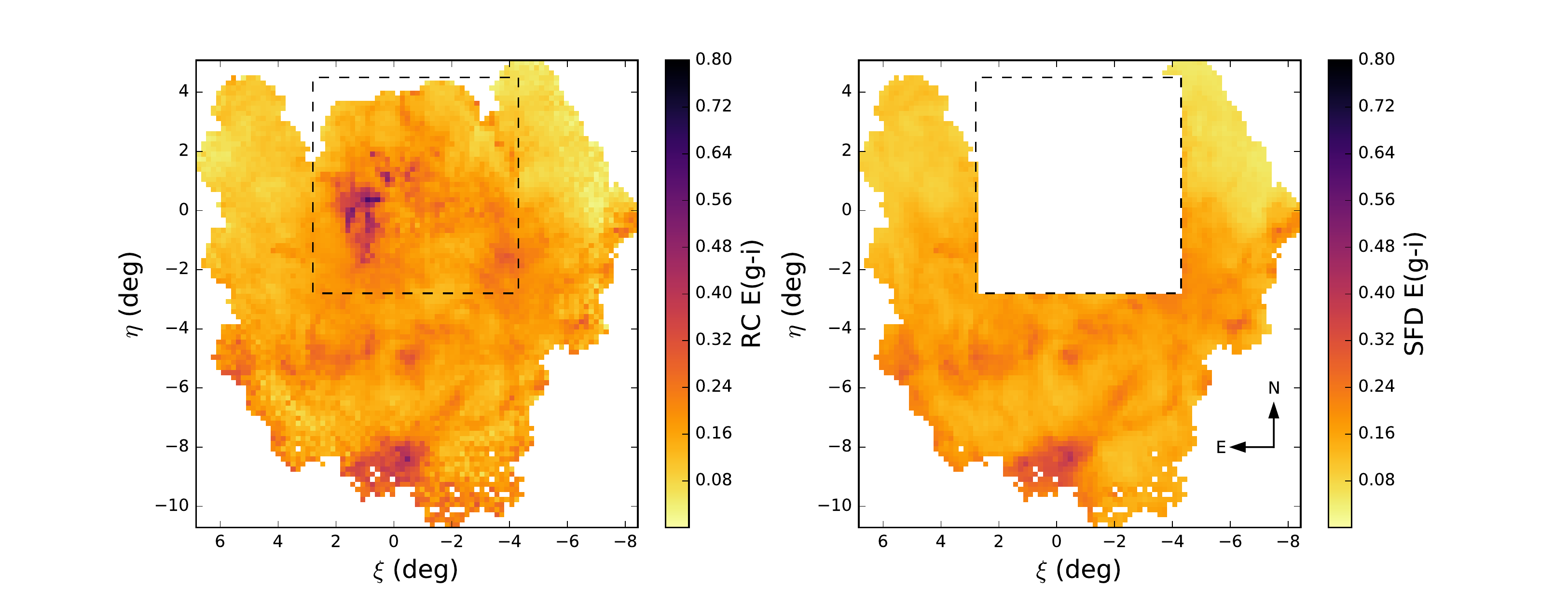}
\caption{$E(g-i)$ reddening maps derived from our RC (left) and from the SFD98 $E(B-V)$ map (right). We use extinction coefficients for the $E(B-V)$ to $E(g-i)$ conversion that account for a systematic $\sim$14\% overestimation in the SFD98 $E(B-V)$ map \citep[see][]{schlafly11,abbott18}. A dashed box shows the inner region where the LMC emission dominates far-IR emission, and thus the SFD98 map cannot provide reliable measurements due to the lack of angular resolution to resolve complex temperature structures. Thus, we intentionally do not show their map inside the box. Outside the box, the two maps shows an excellent agreement, both qualitatively as well as quantitatively, indicating that the majority of the extinction in the outer disk comes from the MW foreground cirrus. The electronic version of our reddening map is available as supplementary material to this paper.
\label{extmap}}
\end{figure*}

\begin{figure*}[t!]
\centering
\includegraphics[width=20cm,trim=4cm 0cm 0cm 0cm, clip=True]{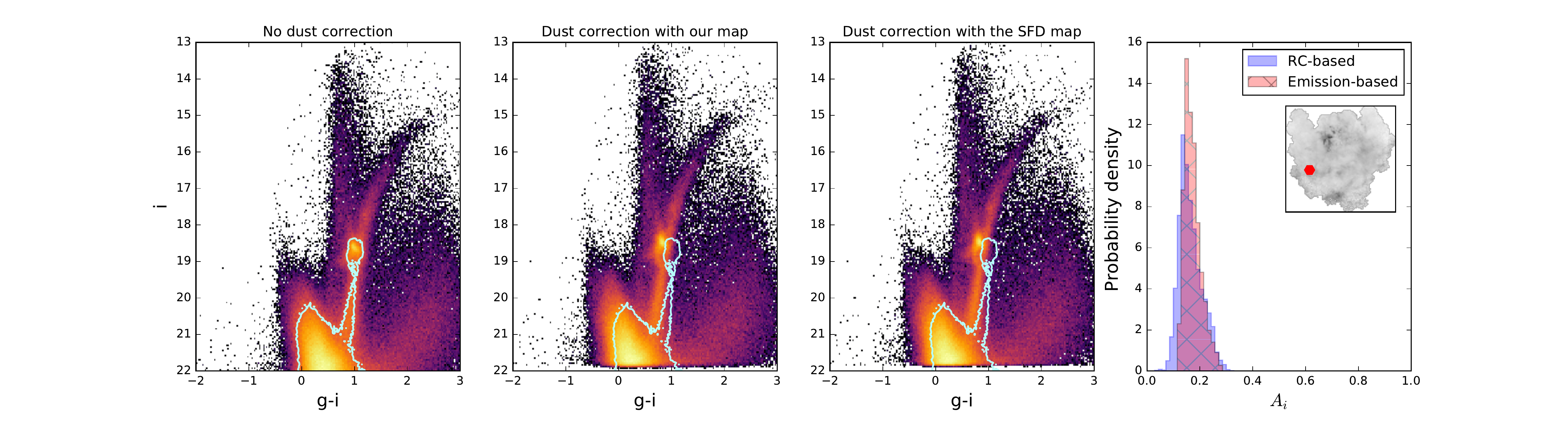}
\includegraphics[width=20cm,trim=4cm 0cm 0cm 0cm, clip=True]{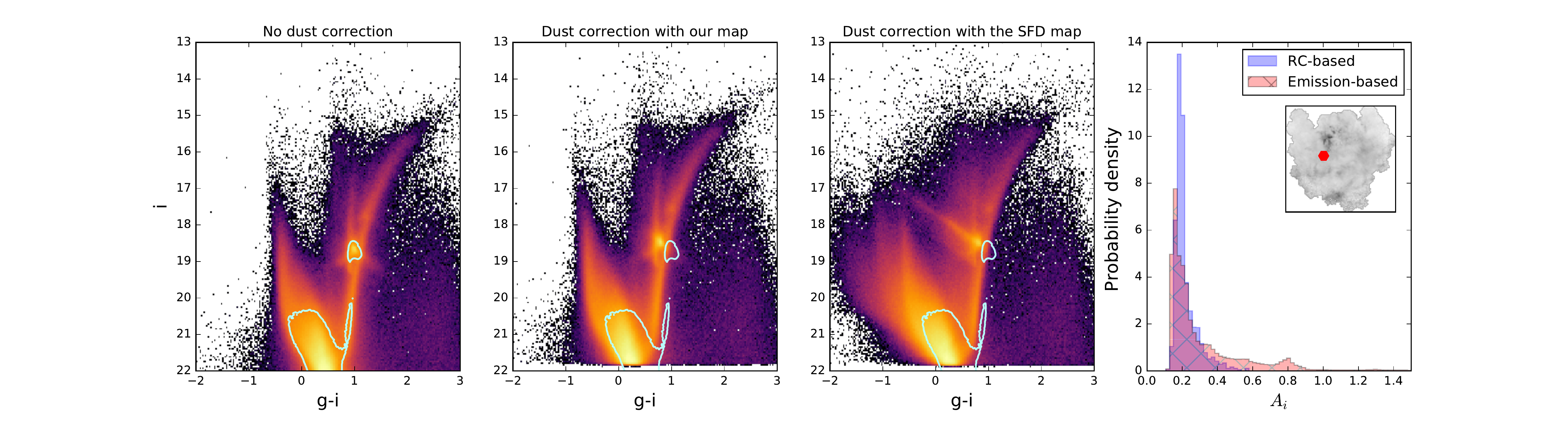}
\includegraphics[width=20cm,trim=4cm 0cm 0cm 0cm, clip=True]{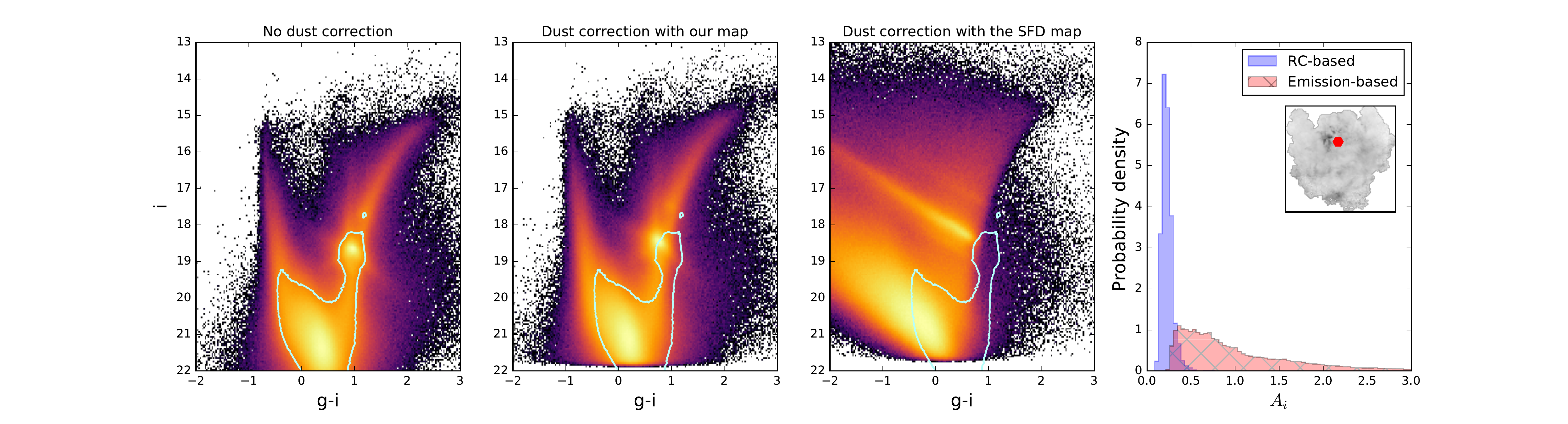}
\caption{Examples for dust correction in low (Field 237), moderate (Field 49), and high (Field 41) extinction regions from top to bottom rows. In each row, we show an observed CMD (left), a CMD corrected using our RC-based reddening map (middle left), and a CMD corrected using the emission-based SFD98 map (middle right). To make comparison easy, we overplot the contours of an observed CMD in the corrected CMDs. In the low-extinction case, both our map and the SFD98 map work well and return almost identical CMDs after correction, confirming that the SFD98 map can be used in low-extinction regions in the outer LMC disk. In the moderate-extinction case, the CMD corrected with our map shows a tighter RGB sequence and a sharper SRC feature while the CMD corrected with the SFD98 map starts suffering from overcorrection, resulting in the RC streak in the opposite direction and artifacts in the MS. As expected, the overcorrection problem gets worse in the high-extinction case, while our reddening map returns the CMD that seems reasonably well corrected. The last columns in each row show the $A_i$ distributions of each field from our and the SFD98 reddening maps. The inset panels show the locations of each field within the LMC (red hexagon). In the low-extinction region, two distributions are very similar to each other. However, in Fields 41 and 49, the distributions of $A_i$ from the SFD98 map feature long tails toward larger values of $A_i$.  As discussed by SFD98 and in the text, the SFD98 maps do not trace extinction well in the inner LMC disk. 
\label{concept_proof}}
\end{figure*}

\begin{figure}
\centering
\includegraphics[width=9cm,trim=0cm 0cm 0cm 0cm, clip=True]{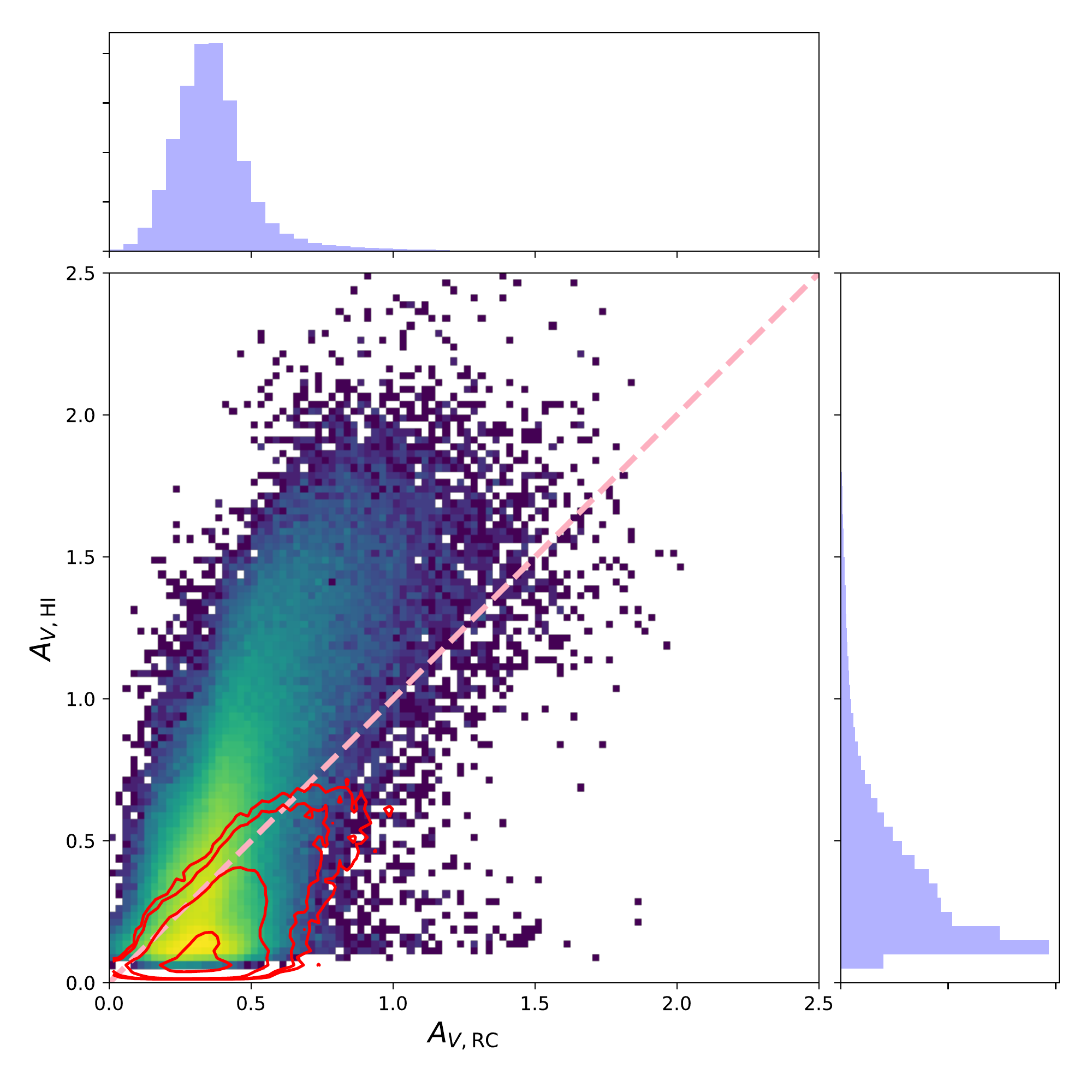}
\caption{A cell-by-cell comparison between the mean $A_{V}$ map from our reddening map and the $A_{V}$ map derived from the \hi\,column density ($N_{\textrm \hi}$) in the inner LMC disk. From $N_{\textrm \hi}$, we determine $A_{V}$ using the average relation found in the LMC within $\sim$8\% uncertainty \citep[see Table 2 of][]{gordon03}. The red contour shows a cell-by-cell comparison with the $A_{V}$ values determined using the relation found around the 30 Dor region \citep{fitzpatrick85}. For both cases, our $A_{V}$ values scale with the $A_{V}$ values derived from $N_{\textrm \hi}$, indicating our RC-based reddening map traces dust in the inner LMC disk reasonably well. The pink dashed line corresponds to a one-to-one relationship. We also present the 1D histograms of the $A_{V}$ values to explicitly show that most of the sightlines have less than 0.5 mag of extinction, indicating that the bias in our RC-based extinction map will likely be no more than a couple of tenths of magnitude.
\label{av_hi_rc}}
\end{figure}

\subsection{Intrinsic Magnitude of the Red Clump}
\citet{girardi01} measured the effect of stellar population variations in the mean RC magnitude for some fields across the LMC disk based on spatially resolved SFHs and AMRs, and reported the maximum difference of 0.036~mag in the $I$-band magnitude among different regions. Using the clean RC sample, we also derive the intrinsic magnitude of the RC after correction for both MW dust and its gradual variation across the inclined LMC disk by assuming that the relative distances (due to the inclined disk) are the primary cause of the magnitude differences observed among the cells \citep{vandermarel01a}. 

First, we compute the median $i$-band magnitude for all 10\arcmin\,cells in the clean RC region after correcting for the extinction by the foreground MW dust. We then adopt the median MW extinction-corrected $i$-band magnitude as a ``fiducial'' RC magnitude to obtain the relative distances to each other. These relative distances are converted to the absolute distances by putting the LMC center at $D_0$ = 49.9~kpc \citep{degrijs14}. We use the absolute distances ($D$) and celestial coordinates (right ascension $\alpha$, declination $\delta$) of each cell to compute their Cartesian coordinates ($X$,$Y$,$Z$). 

To perform the coordinate transformation from the celestial coordinates and distance to Cartesian coordinates, we adopt the LMC disk's photometric center, ($\alpha_0$,$\delta_0$) = (82.25\textdegree{},--69.5\textdegree{}), as the origin \citep{vandermarel01b,vandermarel14}.

Cartesian coordinates ($X$,$Y$,$Z$) with the origin ($\alpha_0$,$\delta_0$) are defined as follows:
\begin{equation}
\begin{aligned}
  X &= -D\sin(\alpha-\alpha_0)\cos(\delta), \\
  Y &= D\sin(\delta)\cos(\delta_0) - D\sin(\delta_0)\cos(\alpha-\alpha_0)\cos(\delta),\\
  Z &= D_0 - D\sin(\delta)\sin(\delta_0) - D\cos(\delta)\cos(\delta_0)\cos(\alpha-\alpha_0), 
\end{aligned}
\end{equation}
where the $X$-axis is antiparallel to the right ascension axis, the $Y$-axis is parallel to the declination axis, and the $Z$-axis is toward the observer.  

We fit a plane to the ($X$,$Y$,$Z$) positions of all clean RC cells by minimizing their distances to the plane using an optimization algorithm ({\tt scipy.optimize.leastsq}). The fitting yields an inclination angle of 27.81\degr$\pm$0.23\degr\,and position angle of 146.37\degr$\pm$0.37\degr. From this fitted plane for the clean RC sample, we compute the expected magnitude gradient across the LMC disk and correct the extinction-corrected $i$-band magnitude for this expected geometric effect. This MW dust and inclination correction yields the intrinsic RC magnitude. We note that the inclination and position angle measured based on only the clean RC sample is not our final measurement (see Section~\ref{sec:i_pa} for the final measurement), but there is no significant change in the resulting intrinsic RC magnitude with our final measurements of the inclination angle ($\sim$26\degr) and position angle ($\sim$149\degr). Between these two sets of the inclination and position angles, the maximum geometric effect in magnitude is found to be $\sim$0.03~mag, and the magnitude differences in the most areas are comparable to our photometric uncertainty.

Because we assume that the relative distance is the primary cause of the relative magnitude difference among cells, the residuals in the inclination-corrected magnitudes might allow us to characterize the ``marginal'' population effects in the intrinsic RC magnitude distribution. In fact, assuming that the inclination is the primary cause is physically reasonable in terms of the expected mild stellar population gradient across the disk. Thus, the residual differences in the magnitudes after correction for first-order inclination should contain information on the effect of variation in stellar populations across the LMC on the RC intrinsic magnitudes.

In Figure~\ref{int_i}, we present the intrinsic RC magnitude distributions of the cells (left) and individual stars (right). Although both distributions roughly follow a Gaussian profile with the same mean magnitude of $\sim$18.47, the distribution of individual stars is slightly skewed to brighter magnitudes. Bright stars contributing to this skewness are confined into the end of the prominent spiral arm, suggesting these bright stars might be younger on average. Since RC brightness responds to stellar age more sensitively than color for a given metallicity, a lack of skewness in the color distribution of individual stars can be understood. Nevertheless, the width ($\sim$0.1~mag) of the individual stars' magnitude distribution is consistent with the theoretical dispersion in magnitude due to local age and metallicity spread \citep[e.g.,][]{girardi01,yanchulova17}. Furthermore, we find an excellent agreement between the width (0.031~mag) of the cells' distribution, which reflects a global variation in stellar population effect across the LMC disk, and the maximum $I$-band magnitude difference of 0.036~mag that is measured among different regions in the LMC \citep{girardi01} . 

Figure~\ref{rad_mag} presents the radial profile of the intrinsic RC magnitude in the $i$-band. In contrast to the radial color dependence, the radial profile of the intrinsic RC magnitude is rather flat across the entire observed radial range with only a slightly brighter ($\sim$0.03~mag) magnitude in the radial bin, 2.7\degr\,$< \rho <$ 4\degr, which is likely due to younger populations toward the central region. This magnitude radial profile shows excellent agreement with the expected magnitudes (marked as yellow stars) from the same age, metallicity models that explain the measured intrinsic RC color radial profile (see Section~\ref{sec:col_rad}). Thus, we adopt the mean magnitude of 18.476 as a constant intrinsic RC magnitude between 4\degr\,$< \rho <$ 8.5\degr, and adopt a slope of 0.019~mag~deg$^{-1}$ between 2.7\degr\,$< \rho <$ 4\degr. As with the color radial profile, we set the magnitudes at $\rho$ = 2.7\degr\,and 8.5\degr\,as the default values in the innermost ($\rho <$ 2.7\degr) and outermost ($\rho <$ 8.5\degr) radial bins, respectively. We will also discuss how the choice of different magnitudes affects the resulting 3D structures in Section~\ref{sec:popeffect}.

\begin{figure*}
\centering
\includegraphics[width=18cm]{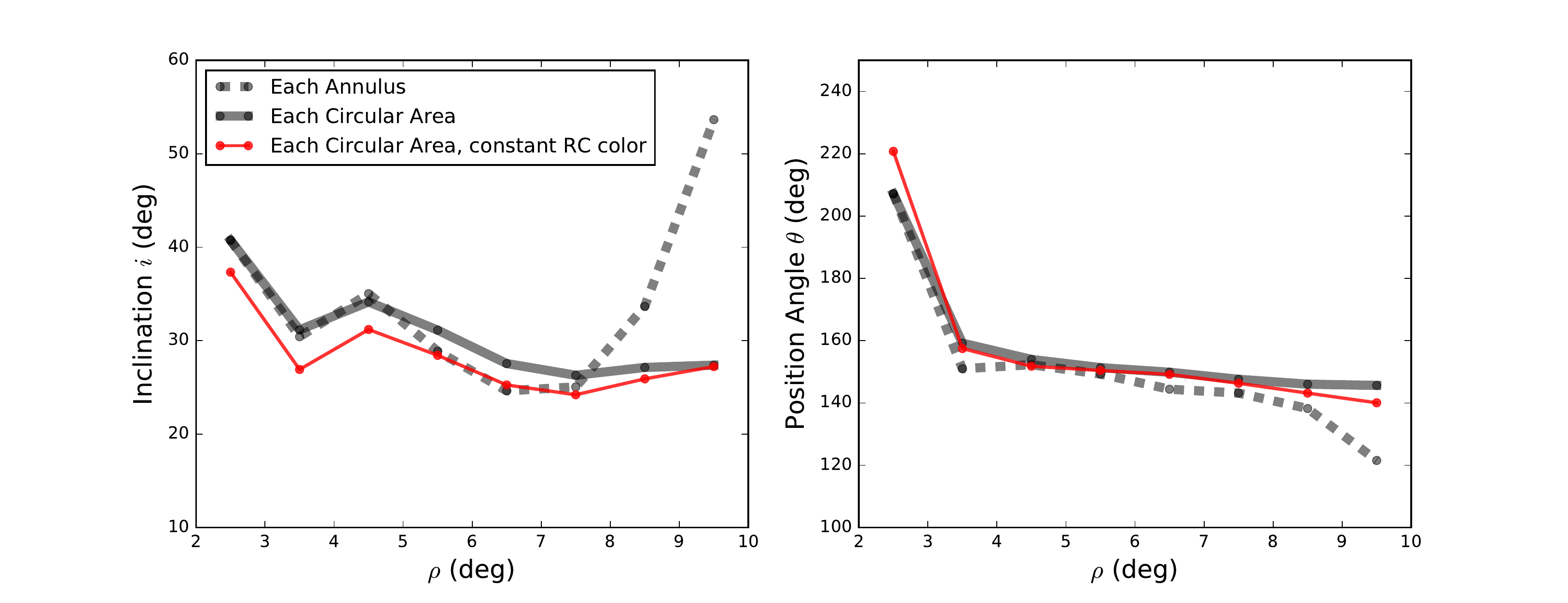}
\caption{Radial dependence of the measured inclination (left) and position angle (right) with our default radial color profile. In each panel, the gray dashed line represents the measurement from each 1\degr\, wide annulus, whereas the gray solid line represents the measurement from each circle with a given radius. The red solid lines show the same as the gray solid lines, but for the case of a constant intrinsic $g-i$ color of 0.820 and a constant intrinsic $i$-band magnitude of 18.476 across the entire disk. In general, both the inclination and position angle decrease with galactic radius, while the annulus fitting returns the increased inclination  and the decreased position angle in the two outer annuli where the outer warp appears.   
\label{incl_pa}}
\end{figure*}


\section{A Reddening Map of the LMC Disk}\label{sec:ext}
Figure~\ref{extmap} presents the reddening map derived by RC color, showing the median reddening along the line of sight toward each cell. We compute the extinction for each cell by comparing the inferred intrinsic color from Section~\ref{sec:int_col} and the observed median color of the RC at a given cell: $E(g-i) = (g-i)_{\rm obs} - (g-i)_{0}$. The average LMC reddening is $E(g-i) = 0.15\pm0.05$, which is in good agreement with the previously reported average reddening measurements using intermediate/old stars in the LMC \citep[e.g.,][]{zaritsky04a,haschke11}. The electronic version of our reddening map is available. 

In the right panel of Figure~\ref{extmap}, we also present the SFD98 reddening map derived from the dust emission. In the region outside the dashed box, our RC reddening map shows excellent agreement with the SFD98 map in both qualitative and quantitative aspects. \citet{sfd98} derived $E(B-V)$ using dust emission at 100$\mu$m assuming only MW dust ($R_V$ = 3.1), and thus we also assume MW dust when converting SFD98 $E(B-V)$ to $E(g-i)$. We note for the reader that \citet{sfd98} reported that for regions containing the LMC, their reddening measurements are unreliable because of insufficient angular resolution to resolve the complex temperature structure toward the LMC disk. The effect of an unresolved temperature structure on the SFD98 map is especially severe in the central regions of the LMC disk where the LMC dust overwhelms the MW foreground. In the central regions, our reddening map shows an excellent morphological agreement with the high-resolution {\it Herschel} dust emission images \citep{gordon14}.    

The distribution of ratios between our reddening map and the SFD98 map can be described as a Gaussian profile with a peak around 1 when ignoring the central region (i.e., high-extinction regions), suggesting good quantitative agreement in the outer disk. This indicates that the majority of the reddening toward the non-central regions results from the MW foreground, which acts as a dust screen on stars behind it, and thus corroborates that the SFD98 map works well in the outer regions. If the LMC internal dust contributes to the majority of the reddening in the outer disk, one should expect discrepancy between the two maps. This is because the SFD98 map traces the total dust emission from a given region, whereas our map represents the median reddening values in each cell measured from individual RC stars that only sample dust in front of them. Thus, the SFD98 map would have shown systematically higher values than our map if internal LMC extinction dominated in the outer disk.  

Figure~\ref{concept_proof} presents examples of CMDs dust-corrected by our map and the SFD98 map for low (Field 237), moderate (Field 49), and high (Field 41) extinction regions. In each case, we show an observed CMD, a corrected CMD using our RC-based reddening map, a CMD corrected using the emission-based SFD98 map, and distributions of $A_i$ values from our and SFD98 reddening maps. For the low-extinction case (top row), both our map and the SFD98 map return almost identical CMDs after correction, and their $A_i$ distributions are indeed very similar to each other, confirming that the SFD98 map can be used in low-extinction regions in the outer LMC disk. For the moderate extinction case, the CMD corrected by our map shows a tighter RGB sequence and a sharper SRC feature, while the CMD corrected by the SFD98 map shows the RC streak in the opposite direction and some artifacts in the MS because of overcorrection. The $A_i$ distribution from the SFD98 map clearly develops a long tail to larger values of $A_i$. As expected, the overcorrection problem with the SFD98 map becomes more severe in the high-extinction case, while our reddening map returns the CMD that seems reasonably well corrected overall. 

As explained by SFD98 and discussed above, the SFD98 map fails in the optically bright LMC disk because the data on which the map was based could not resolve the temperature structure of the LMC disk. This failure explains the severe overcorrection of extinction in Fields 41 and 49 in Figure~\ref{concept_proof}.  An additional issue is that SFD98 measured the total emission by dust, both foreground and internal to the LMC, along the line of sight, whereas our maps are sensitive to the total foreground extinction but only to the median of the LMC internal extinction.
Thus, our map will either over- or undercorrect for the true extinction, depending on the relative geometry of individual stars and dust.  This bias is likely to be most pronounced in the regions of high extinction found in the central LMC disk ($\rho <$ 2\degr). The extent of this bias will likely be no more than a few of tenths of magnitude as will be seen later in Figure~\ref{av_hi_rc}. In regions of low to moderate internal LMC extinction (i.e., $\rho >$ 2\degr), as shown in Figure~\ref{concept_proof}, the bias should be negligible, as the Galactic foreground extinction makes up a larger fraction of the total extinction in these areas. Furthermore, the median relative distances between LMC subregions, corrected for dust using our map, would not significantly change in spite of the bias. Our use of the RC-derived extinction map is thus unlikely to affect the correct geometry of the galaxy midplane, and thus our results on the LMCÕs three-dimensional structure should be robust against this extinction bias.

In Figure~\ref{av_hi_rc}, we perform a cell-by-cell comparison of our median $A_{V}$ map and the $A_{V}$ map derived from the \hi\,column density \citep{staveleysmith03} for a better quantitative verification of our reddening map in the inner star forming disk where the internal LMC dust is dominant. The \hi\,image used here was convolved to have a spatial resolution of 40\arcsec\,and removed a large-scale structured background emission by \citet{gordon14}. We also construct our RC-based $A_{V}$ map at 40\arcsec\,resolution for a cell-by-cell comparison. For a given \hi\,column density ($N_{\textrm \hi}$), we obtain the optical extinction along a given slightline using the empirical relation obtained by correlating the reddening of stars with the \hi\,column density in the LMC galaxy. A coefficient between the gas column density and extinction varies with the ISM properties (e.g., metallicity and ISM phase). For example, \citet{fitzpatrick85} found the relation of $A_{V} = N_{\textrm \hi}/(7.74\times\,10^{21})$ around the 30 Dor region. On the other hand, \citet{gordon03} measured the average LMC $A_{V} = N_{\textrm \hi}/(3.25\times10^{21})$, implying overall lower gas-to-dust ratios than the 30 Dor region. 

In Figure~\ref{av_hi_rc}, the underlying color scale shows the comparison RC-based $A_{V}$ values with $A_{V}$ values derived from the average LMC relation, while the red contour shows the comparison with $A_{V}$ values derived from the 30 Dor region relation. For both cases, our RC-based $A_{V}$ measurement correlates well with the \hi\,-based $A_{V}$ measurement. This, combined with morphological agreement with the FIR dust emission images, suggests that our RC method works reasonably well in the inner LMC disk as well. Although these two specific coefficients do not result in one-to-one correlation, it seems that the LMC might have a typical coefficient somewhere between these two coefficients, but determination of that coefficient or gas-to-dust ratio is beyond the scope of this paper. Besides the variation in gas-to-dust ratio across the LMC star-forming disk, another obvious thing that might contribute to the deviation from one-to-one correlation is the bias in our RC-based $A_{V}$ values especially at larger values of $A_{V}$ where the impact of difference between our median extinction and the total line-of-sight extinction toward individual stars is more significant. For the average LMC case, lower gas-to-dust ratios result in larger extinction values for a given \hi\,column density than our RC-based median extinction values. On the other hand, for the 30 Dor case, our RC-based $A_{V}$ values are larger because larger gas-to-dust ratios lead to lower $A_{V}$ values for a given \hi\,column density. These trends, shown in Figure~\ref{av_hi_rc}, may in part be explained by the bias in our RC extinction map. The 1D histograms of $A_{V}$ values in Figure~\ref{av_hi_rc} show that most line of sights have extinction less than 0.5~mag, indicating that the bias in our RC-based extinction map for the central star-forming disk is likely no more than a few tenths of magnitude.

\section{A Three-dimensional Map of the LMC Disk}\label{sec:3Dmap}

\subsection{Inclination and Position Angle}\label{sec:i_pa}
In this section, we constrain the LMC disk's ($i$,$\theta$) using RC stars selected from the data covering a wide area of the unexplored southern part of the disk. First, we measure relative distances with respect to the LMC center using the extinction-corrected relative $i$-band magnitudes for each subregion. With a distance to each subregion, the subregion positions are expressed in Cartesian coordinates ($X$,$Y$,$Z$). 

To characterize the properties of the LMC disk, we define a galaxy plane relative to the sky plane based on the 3D distribution of all cells by minimizing their distances to the plane using {\tt scipy.optimize.leastsq} again. The unit normal vector ($\hat{n}$), defined in the Cartesian coordinates, of the best-fit plane is ($n_x,n_y,n_z$) = (0.375,-0.223,0.899). From this normal vector, we derive $i = \cos^{-1}(n_z)$ and $\theta = \pi + \tan^{-1}(n_y/n_x)$. The fitted plane has an inclination of 25.86\degr$\pm$0.19 and the position angle ($\theta$) of 149.23\degr$\pm$0.49. The uncertainties reported here are random errors, associated with the plane fitting, calculated using the MCMC sampler {\tt emcee} \citep{foremanmackey13}. We allow each of the 100 walkers to take 2000 steps and compute the 68\% confidence interval as the 1$\sigma$ uncertainty of each fitting quantity after discarding the burn-in phase of the first 500 steps. These errors do not include any systematic uncertainties associated with stellar population effects. The systematic uncertainties in our ($i$,$\theta$) measurements are provided in Section~\ref{sec:discussion}. 

The LMC disk has been found to be twisted (varying line of nodes with galactic radius), warped (not a single plane), and flared (increasing scale height with galactic radius) likely due to tidal interactions with both the SMC and the MW \citep{vandermarel01a,olsen02,nikolaev04,subramanian09,balbinot15}. Therefore, the inclination ($i$) and the position angle of the line of nodes ($\theta$, the intersection of the galaxy plane and the sky plane) depend on the spatial coverage of the survey data. Indeed, it has been shown that these two parameters vary with both angular distance from the galactic center and measurement techniques even for a given stellar tracer because of the complicated shape of the LMC disk \citep{vandermarel01a,subramanian13,jacyszynDobrzeniecka17}. Furthermore, it has been found that determined the LMC disk properties depend significantly on the choice of stellar populations to trace the disk structure \citep[e.g.,][]{balbinot15}. 

In Figure~\ref{incl_pa}, we explore the effect of twists and warps on ($i$,$\theta$) measurement. We first define eight 1\degr\,wide radial annuli between 2\degr\,and 10\degr\, on the tangent plane ($\xi$,$\eta$), and then (1) measure ($i$,$\theta$) from all RC cells in each annulus and (2) measure ($i$,$\theta$) for all RC cells enclosed within a circular area that defines the outer edge of each annulus. Each annulus fitting reflects the local geometry better than the circular area fitting. Black dashed lines present how the ($i$,$\theta$) measurements change for each annulus, while black solid lines show the dependence of ($i$,$\theta$) on areal coverage (i.e., each circular area for a given angular distance from the LMC center). Like \citet{vandermarel01a}, we also find a strong radial dependence of ($i$,$\theta$) -- in general, both ($i$,$\theta$) rapidly (slowly) decrease with galactic radius in the inner (outer) disk. The inclination and position angle are large in the central region ($<$ 4\degr), where the bar is dominant. This indicates that the bar itself is likely tilted more than the rest of the disk (see Section~\ref{sec:warpbar}). \citet{inno16} studied the 3D distribution of Cepheids and also found a strong radial dependence of ($i$,$\theta$) in the inner LMC disk. The inclination angle from annulus fitting significantly increases in the very outer region ($>$ 8\degr), where the warp is dominant. 

Both measuring methods (annulus vs.\ circular area) produce similar radial trends, except for the outer region ($\rho >$ 8\degr). The inferred inclination angle dramatically decreases from $\sim$40\degr\,to $\sim$25\degr\,with angular distance until $\rho \simeq$ 7\degr\,and then almost converges or slightly increases beyond that radius. On the other hand, the $\theta$ values are rather flat with a slight decrease after a sharp drop at $\rho <$ 3.5--4.5\degr, ranging between $\sim$140\degr\,and $\sim$160\degr. This trend is found in both measuring schemes, although the measured $\theta$ values from the annulus fitting are slightly smaller, and there is a drop to $\sim$120\degr\,in the outer region. 

\begin{figure*}
\centering
\includegraphics[trim=2cm 0cm 0cm 0cm, clip=True, width=20cm]{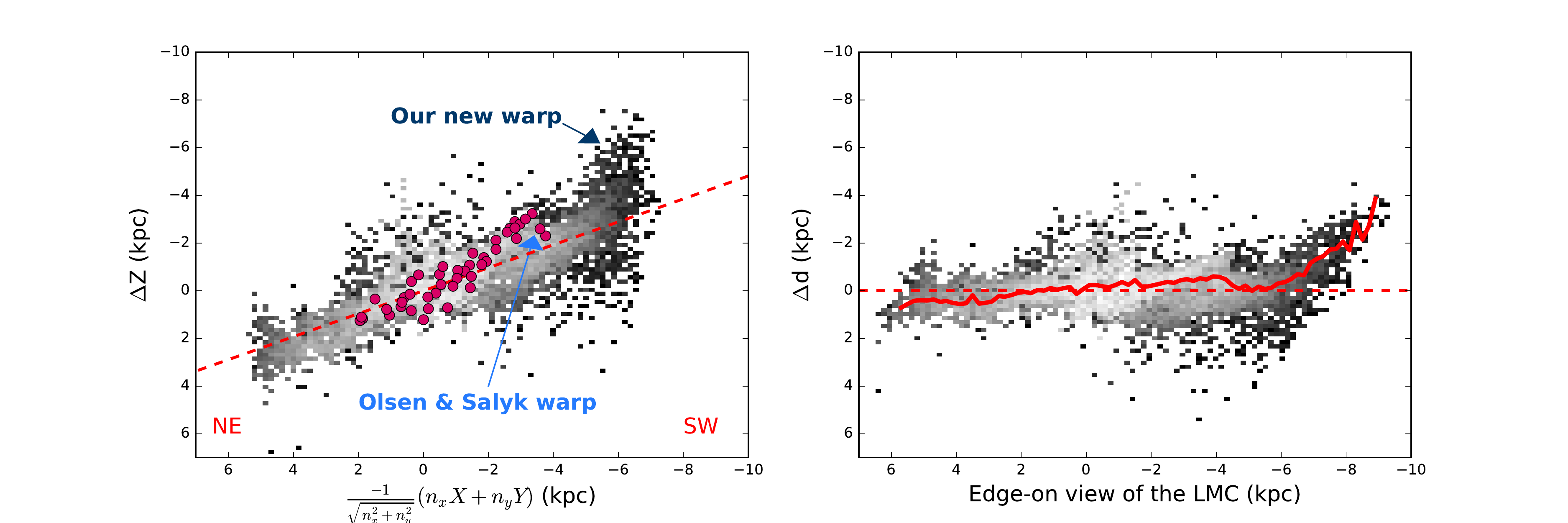}
\caption{{\it Left:} The 3D RC distribution along the maximum line-of-sight depth gradient, which is perpendicular to the line of nodes. Positive (negative) $\Delta\,Z$ denotes closer (farther) to the observer than the galaxy center. The grayscale represents the star number counts -- the brighter the pixel, the higher the stellar number density. The LMC disk is clearly tilted with respect to the sky plane ($Z$ = 0) in a way that the NE is closer to us and the SW is farther away from us. We mark the warp found in the inner part of the disk by \citet{olsen02} as well as the new warp in the outer disk toward the SMC. For clarity, we present the cells corresponding to the \citet{olsen02} observing fields as red circles. Out of the 50 observing fields of \citet{olsen02}, 44 fields overlap with our data. {\it Right:} An edge-on view of the LMC disk. The y-axis is the distance from the fitted disk in this projection. The red solid line traces the median distance weighted by stellar number density. 
\label{3d}}
\end{figure*}
We also investigate the radial dependence of ($i$,$\theta$) values for the case in which a constant intrinsic RC color of 0.82 is used across the disk. The red lines in Figure~\ref{incl_pa} show ($i$,$\theta$) results for the case of the constant intrinsic RC color and circular area fitting. The inclination angles behave in an almost identical way to those from the radially varying intrinsic RC color case in all different areal coverages. There are only a few degree differences along the galactic radius. The position angles for the constant intrinsic RC color case also behave in a similar way to the radially varying intrinsic RC color case. For the ($i$,$\theta$) measurements in each annulus, the constant color case follows almost the same pattern as the radially varying color case, and thus we do not overplot them for clarity. 

For the LMC, its previously reported inclination angle ranges from $\sim$7\degr\,to $\sim$40\degr,and its reported position angle also widely ranges from $\sim$100\degr\,to $\sim$180\degr\,\citep[e.g.,][]{vandermarel01a,olsen02,koerwer09,subramanian10,subramanian13,inno16,jacyszynDobrzeniecka17}. Our measured $i$ and $\theta$ values, both from the best-fit plane and from the annulus/circular area fitting, fall well within these ranges. Specifically, the measured position angle is in good agreement with the reported values based on the RC \citep[e.g.,][]{olsen02}. On the other hand, the measured $i$ is smaller than reported values based on the RC. This discrepancy from face-value comparison is likely due to our larger areal coverage of the southern disk than previous studies. If we consider the smaller areal coverage of the different datasets based on Figure~\ref{incl_pa}, the measured $i$ agrees well with the previous measurements. Our inclination is also similar to that measured in the northern outer disk by modeling the star number count map as an elliptical disk \citep[$i$ = 25.18\degr$\pm$0.71;][]{mackey16}. Furthermore, our results are consistent with the measured ($i$,$\theta$) based on combined information of line-of-sight velocities and proper motions \citep{vandermarel14}. They measured ($i$,$\theta$) = (30.3\degr$\pm$5.9,153.7\degr$\pm$5.4) for their young star sample (red super giant stars) and (34.0\degr$\pm$7.0,139.1\degr$\pm$4.1) for their old star sample (a mix of carbon stars, AGB stars, and RGB stars).

\subsection{Disk Thickness}
We measure the thickness of the LMC's stellar disk from the distribution of distances of individual stars to the fitted plane. The distribution of distances to the fitted disk plane roughly follows a Gaussian distribution with a $\sigma$ of $\sim$1.0~kpc, leading to the FWHM thickness of $\sim$2.35~kpc. The LMC disk has been known to be thicker than the MW thick disk. 

The line-of-sight depth of the LMC disk has been measured in the literature \citep[e.g.,][]{subramanian09,jacyszynDobrzeniecka17,yanchulova17}. We also measure a line-of-sight depth by looking at the distribution of individual stars along the $Z$-axis. We use stars only at $\rho >$ 2\degr\,to obtain a robust measurement by avoiding the artificial broadening along the line of sight due to the bias in our dust correction from using the median reddening values (see bottom panels in Figure~\ref{concept_proof}). The measured line-of-sight depth is $\sim$7~kpc, which is larger than previous measurements by 1--2~kpc. This larger depth might be due to our larger areal coverage with the outer warp. As expected, including stars in the central region ($\rho <$ 2\degr) increases the line-of-sight depth by $\sim$0.5~kpc.

\subsection{Warp and Bar}\label{sec:warpbar}
The left panel of Figure~\ref{3d} shows the 3D structure of the LMC disk in projected Cartesian coordinates ($X$,$Y$,$Z$) along the maximum distance gradient, which is perpendicular to the line of nodes. The grayscale reflects the stellar number counts along the viewing direction. The LMC disk is tilted against the sky plane, with the northeast (southwest) closer to (farther from) us. As shown, the LMC disk is clearly not a simple plane, which is consistent with previous studies -- it has multiple ripples like the MW or other spiral galaxies \citep[e.g.,][]{gomez13}. Similar ripples in the inner region can be also found in the near-infrared photometric data of the RC \citep[Fig.9 in][]{subramanian13}. A rippled disk is a common feature in a galaxy with close encounters with its satellite galaxies \citep{gomez17}.

In the LMC inner disk, \citet{olsen02} detected a warp feature in the southwest (between 2\degr\,and 4\degr\,from the center along the maximum gradient). We reproduce their warp in our data as well by looking at cells corresponding to their observing fields (red circles Figure~\ref{3d}). These cells coincide with the highest stellar density. We mark their warp the ``Olsen \& Salyk warp" in Figure~\ref{3d}, which turns out to be a portion of the rippled disk. Its amplitude ($\sim$2~kpc) and direction (toward us) are in excellent agreement with the findings in \citet{olsen02}. The Olsen \& Salyk warp corresponds to the feature seen between --2.5~kpc and --4~kpc along the maximum gradient. Its direction toward us makes the warp being tilted opposite to the rest of the southwest disk. \citet{vandermarel01a} suggested a warped disk based on decreasing inclination angle with angular radius between 2.7\degr\,and 7\degr, indicating that the Olsen \& Salyk warp indeed affected their inclination and position angle measurements. We also detect such a decrease in inclination until $\sim$7\degr\,from the center (see Figure~\ref{incl_pa}). When measuring the disk inclination using the cells in the \citet{olsen02} observing fields, but excluding the cells that form the Olsen \& Salyk warp, the fitting yields $i = $34.42\degr\,which is consistent with $i =$ 35.8\degr$\pm$2.4\degr\,measured by \citet{olsen02}.  

In the right panel of Figure~\ref{3d}, we present the edge-on view of the disk. The red line shows the average distance to the fitted plane weighted by stellar number count in each 0.2~kpc bin along the x-axis. With our large areal coverage for the southern disk, we find a new prominent warp in the southwest that starts at about --5.5~kpc from the center (or $\sim$7\degr\,southwest from the center on the sky). The warp departs from the fitted disk plane up to $\sim$4~kpc in a direction away from the Sun, which is in the opposite direction to the Olsen \& Salyk warp. This is the first detection of this outer warp, made possible by the large areal coverage of the SMASH data for the southern LMC disk. The warp feature is robust against the choice of the intrinsic RC color radial profile. Also, we do believe that the warp is not an artifact resulting from the edge of our observing area, which is imprinted as a sharp drop in stellar number density that is coincidentally parallel to the direction of the new warp. This is because the average distance from the galaxy plane starts to increase negatively (i.e., away from the observer) between --5~kpc and --6.5~kpc where the effect of the observing edge does not kick in yet.     

\begin{figure}
\centering
\includegraphics[width=9cm,trim=0cm 0cm 0cm 1cm, clip=True]{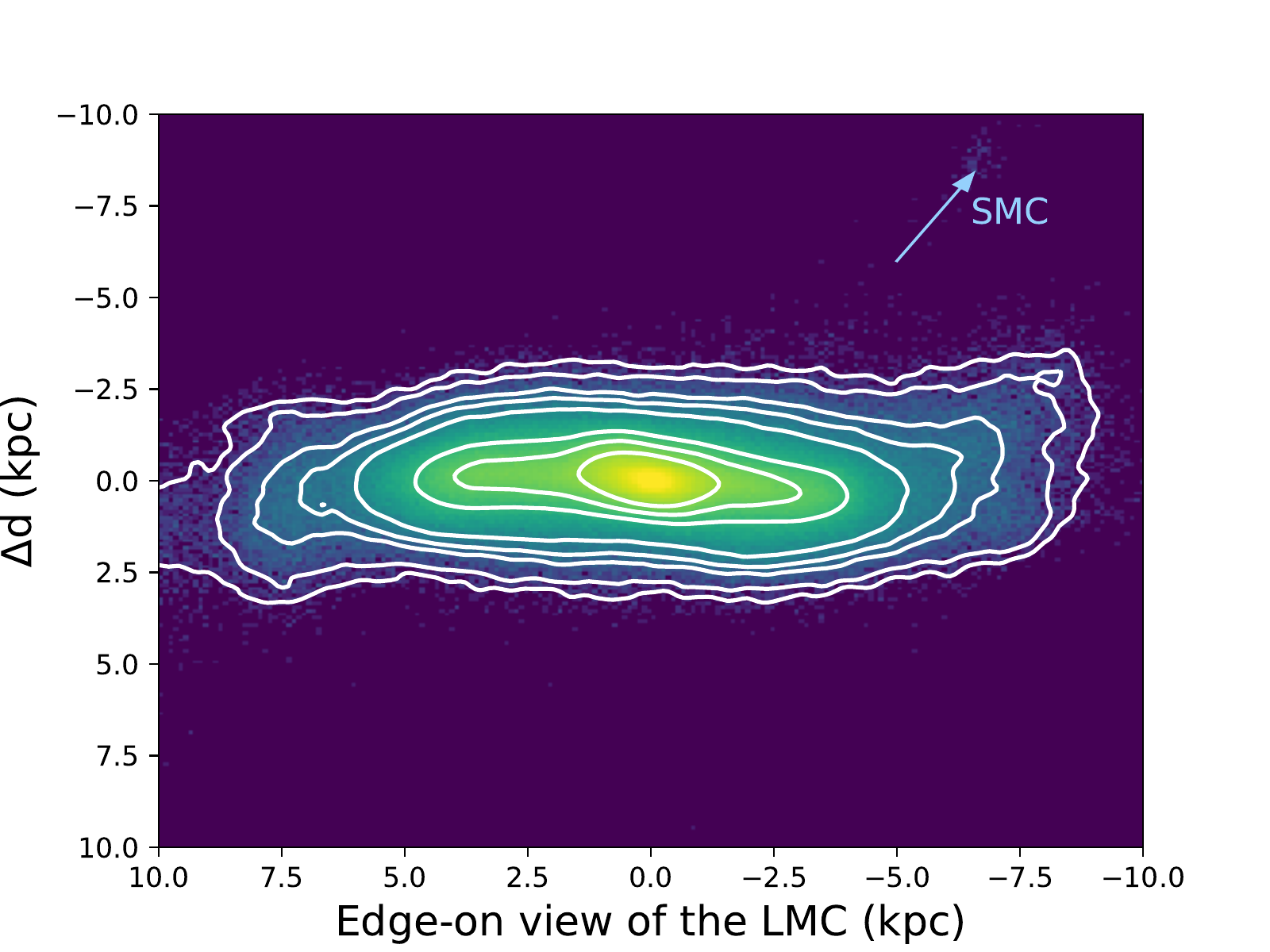}
\caption{A modified version of the edge-on view of a stellar density map for Model 2 from \citet{besla12}. This projection is comparable with our edge-on view shown in Figure~\ref{3d}. This model predicted the outer warp we detect for the first time in the outer disk toward the SMC due to the recent direct collision between the MCs a few hundred Myr ago. The direction toward the SMC is marked.
\label{besla12_model2}}
\end{figure}

The bar stands out as a high density in the central region in Figure~\ref{starcount}. The 3D spatial distribution of cells around the bar -- it is non-trivial to identify only bar stars -- shows a wide dispersion along the vertical direction (i.e., perpendicular to the plane). Thus, quantifying the bar structure as a plane is not reasonable. Instead, we can interpret the larger ($i$,$\theta$) of the inner disk ($\rho <$ 4\degr), which encompasses the bar, as evidence of a highly tilted bar. If the central bar significantly affects the determination of the geometry of the inner disk, the bar seems to be tilted relative to the fitted plane at least by $\sim$5--15\degr. Although the origin and the geometry of the off-centered bar in the LMC remains to be fully understood \citep[e.g.,][]{zhao00,zaritsky04b,bekki09,subramaniam09}, an off-centered and tilted bar naturally forms in Besla's Model 2 due to a recent direct collision with the SMC.

\section{Discussion} \label{sec:discussion}
\subsection{Implication for the LMC--SMC Interaction Histories}
The SMASH data enable us to constrain the 3D structure of the LMC disk and detect its interesting features. The LMC disk is found to be twisted and warped with ripples, and to have a tilted bar. These features were previously found in other studies, and are mainly associated with the inner LMC disk. With our SMASH data, which includes a large unexplored area of the southern disk, we also detect these features, confirming that the LMC is indeed significantly disturbed. Furthermore, we reveal a prominent warp in the outer southwest disk ($\rho >$ 7\degr) for the first time. This warp departs $\sim$4~kpc from the LMC disk plane toward the SMC.   

The current space velocities of the MCs favor their first infall into the MW halo, suggesting interactions with the SMC as the primary cause of the disturbed LMC disk rather than MW tides. Thus, morphological studies of the MCs are one of the keys to constraining their interaction history. Under the first infall scenario, \citet{besla12} investigated the role of tidal interactions between the MCs in their evolution using two different orbital histories between the MCs: (1) Model 1 -- the SMC completes two passages around the LMC without getting closer than 20~kpc, and (2) Model 2 -- the SMC completes three passages around the LMC including the recent ($\sim$100-300~Myr ago) direct collision. Each model successfully reproduces some properties of the Magellanic system, and while Model 1 better reproduces the large-scale gaseous structures, Model 2 performs significantly better at reproducing the detailed morphology and kinematics of the LMC disk. 

The edge-on view in Figure~\ref{3d} is comparable to the galaxy y--z plane shown in the bottom panels of Figs. 12 and 13 from \citet{besla12}. For convenience, we show a modified version of an edge-on view stellar density map for Model 2 from \citet{besla12} in Figure~\ref{besla12_model2}. Their Model 2 clearly shows the outer warps in both the stellar and gaseous disks. 

Although \citet{besla12} tried to connect their outer warp with the Olsen \& Salyk warp, which was the only prominent observed warp at that time, our study makes it clear now that the Olsen \& Salyk warp is one of the inner disk features. Thus, their predicted outer warp, starting at $\sim$5~kpc from the disk center in their Model 2, actually has not had an observed counterpart until now. Based on the very similar position and size of the warp we find in the outer southwest disk, we argue that our new warp is the one corresponding to the outer warp presented in Model 2. Indeed, the position and amplitude of the warp is surprisingly similar to those predicted in their Model 2.     

By contrast, the big warp in the outer disk does not appear in their Model 1, in which the SMC completed only two passages with large separations around the LMC. Because this outer warp is not created in Model 1, which has no direct collision between the MCs, our finding of the new warp (in addition to the tilted bar) supports a recent direct collision of the LMC with the SMC. The observed relative velocities of the SMC and LMC measured using new HST proper motions of the SMC \citep{zivick18} further support this direct collision scenario. 

More importantly, this new morphological feature might provide better constraints on model parameters for future theoretical modeling, not just for the MCs themselves but possibly also for other dwarf--dwarf pair systems \citep[e.g.,][]{stierwalt15,pearson16}. To provide a more complete picture of the outer warp, future studies should attempt to identify and quantify its counterpart in the northeast disk. If the warp originates owing to the tidal field of the MW, we expect an ``integral'' shape \citep[e.g.,][]{mackey16,gomez17}. If the structure originates from a collision with the SMC, the warp in the northeast is expected to be weaker than the outer warp we find \citep{besla16}.

\subsection{Effect of Variation in Stellar Populations}\label{sec:popeffect}
Although the RC is a good probe of extinction and distance, it suffers from stellar population effects in its color and magnitude. In the previous section, we measured the radial profiles of the intrinsic color and magnitude using only stars from the clean RC regions across the LMC disk to take population variations into account as much as possible. This analysis uncovered a radial variation both in color and magnitude. However, the limited areal coverage (2.7\degr\,$< \rho <$ 8.5\degr) of the clean RC regions prevents us from obtaining complete color and magnitude radial profiles that fully cover from the innermost region ($\rho <$ 2.7\degr) to the outer region ($\rho >$ 8.5\degr). 

The ($i$,$\theta$) and amplitude of the warp might depend on our choice of the intrinsic RC color in the innermost and outermost radial bins. The amount of the extinction correction is determined by the choice of the intrinsic colors, and a larger extinction correction to the observed magnitude leads to a brighter intrinsic magnitude (i.e., closer to us), and vice versa. If we adopt a redder (bluer) intrinsic color for the outermost radial bin, then the distances to the regions associated with the warp feature would become larger (smaller), making the warp more (less) prominent. Furthermore, variations in the intrinsic magnitude in the innermost and outermost radial bins also affect the relative shape of the disk, probably leading to a change in the warp amplitude.    

In this section, we attempt to adopt reasonable extremes of the assumed intrinsic color and magnitude for those regions based on both our own measurements in the rest of the radial range and the results of previous studies on the LMC's metallicity gradient and AMR. We discuss how our choice of the intrinsic color and magnitude in the central and outer regions can affect the 3D structural measurements. We also discuss the case for a single intrinsic color and magnitude across the disk to compare with the traditional RC method.

{\it Case I:} For the innermost radial bin ($\rho <$ 2.7\degr), if we adopt \mbox{[Fe/H]} = --0.4 and a $\sim$1~Gyr old\footnote{This is the young end for the mean RC age range \citep{girardi16}.} population, the predicted mean RC color is $\sim$0.84, which is redder than the rest of the disk, and the corresponding mean RC magnitude is $\sim$18.76~mag, which is $\sim$0.3~mag fainter than the default magnitude in the innermost radial bin. 

This fainter intrinsic magnitude in the innermost region induces a structure protruding from the disk by 4--5~kpc, making the outer warp slightly less prominent than the default case (by $\sim$10\%). For the main disk (i.e., excluding the protruding structure), we obtain the inclination of 26.56\degr\,and position angle of 147.02\degr\,with respect to the results for our default case. 

An elevated bar above the disk in the LMC has been suggested by some studies \citep[e.g.,][]{zhao00,nikolaev04,haschke12b}. For example, \citet{haschke12b} found evidence that the bar is $\sim$5~kpc closer to us than the disk in their RR Lyrae sample, but no evidence was found in their Cepheid sample. On the other hand, \citet{jacyszynDobrzeniecka17} interpreted the protruding structure seen in their OGLE-IV RR Lyrae stars as an artifact due to blending and crowding effects. In addition, \citet{subramaniam09} analyzed the OGLE-III RC stars and concluded that the bar is located within the disk. We also do not take this protruding structure seriously because the inferred reddening from this color choice is lower than the previous measurements for the central region \citep[e.g.,][]{haschke11,tatton13}.    

Although it might be reasonable to adopt this combination of age and metallicity for the innermost radial bin, the intrinsic RC color and magnitude vary significantly in the young age regime \citep[$<$ 1.5~Gyr; see Fig. 1 in][]{girardi01}. This means that it is impossible to determine representative ages from our own clean RC sample without detailed SFH measurements in the central region. Stellar populations in the central region might be too complex to be described as one representative color and magnitude.

{\it Case II:} For the outermost radial bin ($\rho >$ 8.5\degr), if we extrapolate with the outer radial bin's slope of --0.033~dex~deg$^{-1}$, the expected color at $\rho =$ 10\degr.5 is $\sim$0.7. This color can be reasonably explained with a slightly lower metallicity (\mbox{[Fe/H]} = --0.9 or --1.0) and older age ($\sim$10~Gyr or $\sim$8~Gyr) compared to the previous radial bin, which is also consistent with the LMC's smooth and shallow metallicity and age gradients. If we adopt the population of \mbox{[Fe/H]} = --0.9 and 10~Gyr, the corresponding model magnitude at $\rho = $10.5\degr is $\sim$18.6~mag, which requires a magnitude profile slope of 0.059~dex~deg$^{-1}$ in the outermost radial bin. With these modified color and magnitude radial profiles, we measure the inclination of 24.47\degr\,and position angle of 155.66\degr. The outer warp becomes more prominent; its  amplitude increases by $\sim$10\% with respect to the results for our default case.

{\it Case III:} If we combine the changes in these two radial bins (i.e., Cases I and II), both the inclination and position angle are consistent with the results for our default case within uncertainties. This might be because the net effect from the bar region and the warp region compensate each other. The warp shows no noticeable change in its shape and amplitude. 

{\it Case IV:} Adopting a constant intrinsic RC color and magnitude is the simplest way to derive line-of-sight extinctions and distances to RC stars. Previously, many studies measured the LMC structure in this way, and then discussed possible uncertainties due to the population effects. If we assume no radial dependence of the intrinsic RC color and magnitude throughout our analysis, the inclination and position angle are 26.59\degr\,($\sim$1\degr\,larger than the default case) and 140.88\degr\,($\sim$9\degr\,smaller than the default case). In this case, the Olsen \& Salyk warp becomes stronger, making the outer warp launch from below the fitted plane. This results in a decrease in an amplitude of the outer warp, which is measured from the fitted plane, by up to $\sim$25\%. If we make a fairer comparison with the other cases where the outer warp launches from the fitted plane by measuring its amplitude from where it actually starts, the amplitude is $\sim$10\% smaller than the default case. 

In conclusion, the systematic uncertainties due to population effects on the inclination are --5.4\%/+2.8\% (corresponding to --1.39\degr\,and 0.73\degr) and --5.6\%/+4.3\% (corresponding to --8.35\degr\,and 6.43\degr) on the position angle. The shape and the presence of the outer warp are robust against the choice of radial profiles of the intrinsic RC color and magnitude. The amplitude of the warp is marginally sensitive (up to $\sim$10\%) to the choice of the intrinsic color and magnitude in the innermost and outermost radial bins. Although there are small variations in the ($i$,$\theta$) measurements with the population effects, the general trends of these parameters do not change: (1) the disk is tilted with respect to the sky plane (the northeast is closer to us, whereas the southwest is farther away from us), and (2) the LMC has a well-developed warp toward the SMC in the southwestern region, and (3) the central bar is tilted relative to the rest of the disk. We also find that the disk thickness is robust against the variation in stellar populations (Cases I--IV) regardless of whether the central disk where our RC-derived median reddening map can cause the bias is included. The line-of-sight depth is also robust against the variation in stellar populations when excluding the protruding structure in the central region.

\section{Summary} \label{sec:summary}
We use the high-quality SMASH data that map $\sim$480~deg$^2$ of the Magellanic System with high precision and accuracy both in photometry and astrometry. Out of 197 SMASH fields, we use 62 fields that cover the main body of the LMC ($\sim$5\degr\, to the north and $\sim$11\degr\, to the south from the galactic center), and select 2.2 million RC stars to map the LMC's dust reddening and to understand its 3D structure. The SMASH data cover the southern periphery of the LMC disk, which has never been previously explored to this depth.  

In the past, the RC has been extensively used to trace stellar structures because of its narrow color and magnitude ranges. However, its inherent dependence on stellar population (age and metallicity) has not been properly considered. For example, many studies used a constant color and magnitude assuming a single (or average) age and metallicity. In this study, we conduct a careful analysis to determine the intrinsic RC color and magnitude across the LMC disk.    

To measure the intrinsic color and magnitude from our data, we first construct a clean RC sample by selecting stars in regions with negligible internal extinction (i.e., the presence of a clear separation between the main RC and the RGB). From this subsample, we measure the radial profile of the intrinsic color for the radial range (2.7\degr\,$< \rho <$ 8.5\degr). The measured RC color radial profile shows a constant color over the middle part of the disk and bluer colors both for the inner and outer disks. Bluer color in the inner disk can be interpreted as the presence of metal-rich young populations, whereas bluer color in the outer disk can be interpreted as the presence of metal-poor old populations. As a default color profile, we adopt a constant color at $\rho =$ 2.7\degr\,and at $\rho =$ 8.5\degr\,for the innermost and the outermost radial bins, respectively. 

We also investigate the radial dependence of the RC intrinsic magnitude using the clean sample after removing the inclination effect from the extinction-corrected observed magnitude. There is no significant radial change in the intrinsic magnitude between 4\degr\,$< \rho <$ 8.5\degr\,---only a slightly brighter magnitude in the inner region (2.7\degr\,$< \rho <$ 4\degr) likely due to younger populations. As a default magnitude profile, we adopt a constant magnitude at $\rho =$ 2.7\degr\,and at $\rho =$ 8.5\degr\,for the innermost and the outermost radial bins, respectively. 

After accounting for these population effects, we derive the reddening map that recovers the detailed features found from dust emission \citep{sfd98} in the outer disk at an exquisite level and provides more reliable measurements in the inner disk. A cell-by-cell comparison shows that our reddening map scales with the inferred reddening map from the \hi\,column density along the sightlines. The inner region where the SFD98 map fails shows high reddening values that likely result from LMC internal dust at a variety of temperatures, while the foreground MW dust seems to dominate most of the low-extinction regions outside this inner region.

We fit the LMC disk using the extinction-corrected $i$-band median magnitude map with 10\arcmin\,by 10\arcmin\,spatial resolution. The measured inclination and position angle with random uncertainties are 25.85\degr$\pm$0.19\degr\,and 149.23\degr$\pm$0.49\degr, respectively. We also estimate the systematic uncertainties in our measurements due to additional population effects, which result from potential deviations from our default color and magnitude radial profile choice. Based on our choice of the intrinsic colors for the innermost and outermost radial bins, the inclination varies from 24.47\degr\,to 26.59\degr\, and the position angle varies from 141\textdegree{} to 153\degr. We also find a significant dependence of the inclination on the areal coverage of the data, which decreases with increasing areal coverage.

Finally, we detect a prominent warp in the southwestern disk that starts at $\rho \simeq$ 7\degr\, with an amplitude of $\sim$4~kpc toward the SMC direction. This warp is detected for the first time, and is different from the warp found in the inner disk by \citet{olsen02}, which we also detected in our data as a portion of the rippled disk. We also find that the cells around the bar show a broad vertical distribution. If we interpret the large inclination and position angles of the inner disk, which encompasses the bar, as a result of a tilted bar, the bar seems to deviate from the rest of the LMC disk by at least $\sim$5--15\degr. Both the warp and the tilted bar are consistent with those predicted in Model 2 from \citet{besla12}. Thus, we suggest that the newly found warp and tilted bar are the product of a recent direct collision between the LMC and the SMC.

\acknowledgements
We are grateful to the referee for providing helpful comments to improve the paper. Y.C. and E.F.B. acknowledge support from NSF grant AST 1655677. M.-R.L.C. acknowledges support from the European Research Council (ERC) under the European Union's Horizon 2020 research and innovation program (grant agreement No. 682115). A.M. acknowledges partial support from CONICYT FONDECYT regular 1181797. B.C.C. acknowledges the support of the Australian Research Council through Discovery project DP150100862. T.D.B. acknowledges support from the European Research Council (ERC StG-335936). D.M.D. acknowledges support by Sonderforschungsbereich (SFB) 881 ``The Milky Way System'' of the German Research Foundation (DFG), particularly through subprojects A2. Based on observations at Cerro Tololo Inter-American Observatory, National Optical Astronomy Observatory (NOAO Prop. IDs: 2013A-0411 and 2013B-0440; PI: Nidever), which is operated by the Association of Universities for Research in Astronomy (AURA) under a cooperative agreement with the National Science Foundation. This project used data obtained with the Dark Energy Camera (DECam), which was constructed by the Dark Energy Survey (DES) collaboration. Funding for the DES Projects has been provided by the U.S. Department of Energy, the U.S. National Science Foundation, the Ministry of Science and Education of Spain, the Science and Technology Facilities Council of the United Kingdom, the Higher Education Funding Council for England, the National Center for Supercomputing Applications at the University of Illinois at Urbana-Champaign, the Kavli Institute of Cosmological Physics at the University of Chicago, Center for Cosmology and Astro-Particle Physics at the Ohio State University, the Mitchell Institute for Fundamental Physics and Astronomy at Texas A\&M University, Financiadora de Estudos e Projetos, Funda\c{c}\~ao Carlos Chagas Filho de Amparo, Financiadora de Estudos e Projetos, Funda\c{c}\~ao Carlos Chagas Filho de Amparo \`a Pesquisa do Estado do Rio de Janeiro, Conselho Nacional de Desenvolvimento Cientifico e Tecnol\'ogico and the Minist\'erio da Ci\^encia, Tecnologia e Inova\c{c}\~ao, the Deutsche Forschungsgemeinschaft and the Collaborating Institutions in the Dark Energy Survey. The Collaborating Institutions are Argonne National Laboratory, the University of California at Santa Cruz, the University of Cambridge, Centro de Investigaciones En\'ergeticas, Medioambientales y Tecnol\'ogicas-Madrid, the University of Chicago, University College London, the DES-Brazil Consortium, the University of Edinburgh, the Eidgen\"ossische Technische Hochschule (ETH) Z\"urich, Fermi National Accelerator Laboratory, the University of Illinois at Urbana-Champaign, the Institut de Ci\`encies de l'Espai (IEEC/CSIC), the Institut de F\'isica d'Altes Energies, Lawrence Berkeley National Laboratory, the Ludwig-Maximilians Universit\"at M\"unchen and the associated Excellence Cluster Universe, the University of Michigan, the National Optical Astronomy Observatory, the University of Nottingham, the Ohio State University, the University of Pennsylvania, the University of Portsmouth, SLAC National Accelerator Laboratory, Stanford University, the University of Sussex, and Texas A\&M University. 

\facility{Blanco (DECam).}

\software{scipy \citep{jones01}, numpy \citep{vanderwalt11}, matplotlib \citep{hunter07}, ipython \citep{PER-GRA:2007}, astropy \citep{astropy18}, and emcee \citep{foremanmackey13}.}

\bibliographystyle{aasjournal}
\bibliography{LMC3d_refs}
\clearpage

\end{document}